\documentclass[twocolumn]{aastex631}

\usepackage{amsmath}
\usepackage[caption=false]{subfig}
\usepackage{graphicx}
\usepackage{xcolor}
\usepackage{txfonts}
\usepackage{rotating}
\usepackage[mathlines]{lineno}
\shorttitle{Dense association of stars in the Galactic Centre}
\shortauthors{Hosseini et al.}
\graphicspath{{./}{figures/}}

\begin{document}

\title{Discovery of a Dense Association of Stars in the Vicinity of the Supermassive Black Hole Sgr~A*}

\author[0000-0002-3004-6208]{S. Elaheh Hosseini}
\affiliation{I.Physikalisches Institut der Universität zu Köln, Zülpicher Str. 77, D-50937 Köln, Germany}
\affiliation{Max-Planck-Institut für Radioastronomie (MPIfR), Auf dem Hügel 69, D-53121 Bonn, Germany}

\author[0000-0001-6049-3132]{Andreas Eckart}
\affiliation{I.Physikalisches Institut der Universität zu Köln, Zülpicher Str. 77, D-50937 Köln, Germany}
\affiliation{Max-Planck-Institut für Radioastronomie (MPIfR), Auf dem Hügel 69, D-53121 Bonn, Germany}

\author[0000-0001-6450-1187]{Michal Zaja\v{c}ek}
\affiliation{Department of Theoretical Physics and Astrophysics, Faculty of Science, Masaryk University, Kotl\'a\v{r}sk\'a 2, 611 37 Brno, Czech Republic}
\affiliation{I.Physikalisches Institut der Universität zu Köln, Zülpicher Str. 77, D-50937 Köln, Germany}

\author[0000-0001-9240-6734]{Silke Britzen}
\affiliation{Max-Planck-Institut für Radioastronomie (MPIfR), Auf dem Hügel 69, D-53121 Bonn, Germany}

\author[0000-0002-4408-0650]{Harshitha K. Bhat}
\affiliation{I.Physikalisches Institut der Universität zu Köln, Zülpicher Str. 77, D-50937 Köln, Germany}
\affiliation{Max-Planck-Institut für Radioastronomie (MPIfR), Auf dem Hügel 69, D-53121 Bonn, Germany}


\author[0000-0002-5760-0459]{Vladim\'{i}r Karas}
\affiliation{Astronomical Institute of the Czech Academy of Sciences, Bo\v{c}n\'{i} II 1401, CZ-141 00 Prague, Czech Republic}



\begin{abstract}

We focus on a sample of 42 sources {\em in the vicinity of\/} the bow-shock source IRS 1W (N-sources), located at the distance of $6.05''$ north-east of the supermassive black hole (SMBH) Sagittarius A* (Sgr~A*), within the radius of $1.35''$. We present the first proper motion measurements of N-sources and find that a larger subset of N-sources (28 sources) exhibit a north-westward flying angle. These sources can be bound by an intermediate mass black hole (IMBH) or the concentration that we observe is due to a disk-like distribution projection along the line of sight. We detect the N-sources in $H$, $K_s$, and $L$' bands. The north-westward flying sources could be a bound collection of stars. We discuss a tentative existence of an IMBH or an inclined disk distribution to explain a significant overdensity of stars. The first scenario of having an IMBH  implies the lower limit of $\sim 10^4~M_\odot$ for the putative IMBH. Our measurements for the first time reveal that the dense association of stars containing IRS 1W is a co-moving group of massive, young stars. This stellar association might be the remnant core of a massive stellar cluster that is currently being tidally stripped as it inspirals towards Sgr~A*. The second scenario suggests that the appearance of the N-sources might be influenced by the projection of a disk-like distribution of younger He-stars and/or dust-enshrouded stars. 

\end{abstract}



\section{Introduction} \label{sec:intro}

The extended radio source Sgr~A consists of three components in terms of different radiative and spatial properties: Sgr A East, which is a non-thermal supernova remnant, Sgr A West which is a thermal region associated with the ionized mini-spiral structure, and the compact variable radio source Sgr A~* associated with the supermassive black hole (SMBH) of $\sim 4\times 10^6\,M_{\odot} $ \citep[and references therein]{Balick&Brown1974, Goss1996_discovery,Eckart&Genzel1996, Ghez1998, 2010RvMP...82.3121G, Eckart2017, GravityCollaboration2019, EHT,2024arXiv240403522G} at the center of the Milky Way. The thermal Sgr A West region or the mini-spiral \citep{Eckers1983, Lo1983} is detected as a distinct filamentary structure in near- and mid-infrared bands \citep{Clenet2004, Viehmann2006, Muzic2007,2022ApJ...929..178B} as well as in the mm- and radio domain \citep{Zhao2009,Zhao2010,2017A&A...603A..68M}. It is composed of three clumpy streamers orbiting around Sgr~A* that consist of dust and ionized and atomic gas with the dust temperature of $\sim 100\,$K and the gas temperature of $\sim 10^4$\,K \citep{Yusef-Zadeh1998,Kunneriath2012}.  


A surprisingly large number of massive young stars with the age of a few million years resides in the Milky Way's innermost half parsec (pc). They form at least one disk-like structure of clockwise orbiting stars (CW) \citep{Genzel2003, Levin&Beloborodov2003}.
 \cite{Paumard2006} also propose the existence of a second disk of counter-clockwise orbiting stars (CCW), which, however, appears to be less populous. The existence of disk-like structures, which can be traced even within the innermost S cluster \citep{2020ApJ...896..100A}, suggests the association of star-formation in the Galactic Center with the accretion disks that formed during the previous enhanced accretion activity of Sgr~A*.


The mechanism which leads to the presence of young stars with the age of $\lesssim 5 {\rm Myr}$ is still not completely understood because there are several factors in the Galactic Center environment that should be inhibiting the star-formation process, such as intense UV radiation and stellar winds, large turbulence, enhanced magnetic field, and the tidal forces from the SMBH \citep{Morris1993}. This problem is known as the ``Paradox of youth" \citep{Ghez2003}, though with the advance of star-formation and dynamical models in galactic nuclei, it has been possible to explain the occurrence of young stars in the Galactic Center, see e.g. \citet{2016LNP...905..205M} for a review.

Three plausible scenarios are generally discussed:
\begin{enumerate}
\item \textit{in-situ} star-formation approach, in which stars form in the gravitationally unstable accretion disk or within an infalling molecular cloud that gets shocked due to collisions with the surrounding gas or is tidally compressed \citep{Morris1993, Levin&Beloborodov2003,Nayakshin2006,Hobbs&Nayakshin2009, Lu2009, Bartko2009, Paumard2006, Jalali2014b, Bartko2010A};
\item \textit{in-spiral} star formation approach which claims that the star cluster first starts forming outside the central parsec, e.g. within the circum-nuclear disk or even further away in the central molecular zone. The presence of an intermediate-mass black hole (IMBH) can shorten the dynamical friction timescale, which needs to be at most a few million years long in order for stars to be still young when they settle around the SMBH \citep{Gerhard2001, McMillan2003, Hansen2003, PortegiesZwart2006, Berukoff&Hansen2006, Rizzuto2020,2021ApJ...923...69P};
\item\textit{rejuvenation} scenario, which proposes that late-type stars have undergone collisions with other stars, a dense accretion disc, or a jet, which caused their colder outer layers to be stripped off and made them appear hotter and younger than they actually would be \citep{1992AIPC..278...21M,Morris1993, Genzel2003,2020ApJ...903..140Z}. In addition, late-type stars aligned with an accretion disk accrete material from it, which also leads to their rejuvenation by gaining fresh hydrogen \citep{2021ApJ...910...94C}.
\end{enumerate}
Apart from the cluster of young stars centered around Sgr~A* within $\sim 0.5\,{\rm pc}$, there is a distinct concentration of infrared sources IRS 13 located at the projected distance of $\sim 3-4''$ ($\sim 0.12-0.16$ pc) south-west of Sgr A~* \citep{2023ApJ...956...70P,2024AJ....167...41D,2024ApJ...970...74P}. IRS 13 has been studied as a suitable candidate for the first two aforementioned scenarios for the origin of young stars in the central parsec. IRS 13 consists of two components --  the northern part IRS 13N and the eastern part IRS 13E. IRS 13N hosts young, infrared-excess sources \citep{ Maillard2004, Eckart2004, Paumard2006, Muzic2008,Eckart2013}. IRS 13E mainly contains massive Wolf- Rayet (WR) stars and the violent supergiant E1(OBI) \citep{Maillard2004,Moultaka2005, Paumard2006}, whereas the rest of the stars are fainter ones, identified as late-type stars and are most likely the background stars \citep{Fritz2010}.

The velocity dispersion of the IRS 13E sources motivated the speculation about the existence of an IMBH to bind the cluster and thus prevent it from tidal disruption at the given distance from Sgr~A*. The required mass of the assumed IMBH was inferred to be about $\sim 10^4 M_{\odot}$ \citep{Maillard2004, Schoedel2005}. The existence of the prominent X-ray source in the IRS 13E cluster reinforced the discussion about the IMBH. On the other hand, \citet{Zhu2020} argue that a colliding wind scenario between E2 and E4 can explain the X-ray spectrum as well as the morphology of IRS 13E and suggests that there is no significant evidence that IRS 13E hosts an IMBH more massive than $\sim 10^3 M_{\odot}$. In general, the occurrence of IMBHs in nuclear star clusters is expected based on dynamical arguments. \citet{Rose2022} propose that dynamical mechanisms operating in galactic nuclei, specifically black hole-star collisions, mass segregation, and relaxation, are particularly effective in the formation of IMBHs of $\lesssim 10^4 M_{\odot}$. Their findings imply that at least one IMBH is likely to exist in the central parsec of the Galaxy. Another channel for the occurrence of IMBHs in galactic nuclei is their infall within massive stellar clusters \citep{1969ApJ...158L.139S,2002ApJ...576..899P,2004cbhg.symp..138R,2022arXiv220205618F}, which directly relates dense stellar associations such as IRS13 with an IMBH. \citet{Rose2022} suggest that the collisions between black holes and stars can contribute to the diffuse X-ray emission in the Galactic Center region. Hence, also a growing black hole inside IRS13E that collides with the surrounding stars could contribute to the X-ray emission of the cluster.


Another apparent concentration of stars in the inner parsec of the Galactic Center is associated with the source IRS 1W that is located within the northern arm of the mini-spiral. The mini-spiral contains mostly hot, ionized gas traced by bright Br$\gamma$ emission. \citet{Vollmer2000} interpret the kinematics of the ionized gas in the northern arm to be Keplerian. IRS~1W has been studied previously as a bow shock source interacting with the material of the northern arm. Its broad-band infrared continuum emission is consistent with the temperature of $\sim 300$\,K \citep{T02,T03,T05} suggesting the presence of warm dust in its bow shock.

\par \citet{Ott1999} used the speckle camera SHARP on the New Technology Telescope (NTT) and showed that IRS 1W is an extended source which could be a young star. Previously it was suggested that it could have formed in the northern arm of the mini-spiral or from the accreted gas and dust in the infalling material out of the northern arm. However, due to the current low inferred mass of the minispiral gas of $\sim 100\,M_{\odot}$ it is unlikely that it formed within the northern arm. \citet{Tanner2005} observed IRS 1W alongside IRS 10W and IRS 21 by W.M. Keck 10 meter and Gemini 8 meter telescopes and identified them as bow-shock sources with central sources as Wolf-Rayet stars. The stellar-wind properties and kinematics of IRS 1W are similar to the properties of the clockwise orbiting He emission-line stars. \citet{Viehmann2006} showed that IRS 1W is very red and has a featureless spectral energy distribution (SED). \citet{Sanchez-Bermudez2014} analyzed the bow-shock orientation alongside the proper motion and confirmed that IRS 1W is a Wolf-Rayet star with a bow shock created by the interaction between the mass-losing star and the surrounding interstellar gas.


The apparent stellar overabundance in the vicinity of IRS 1W could hypothetically be a stellar cluster similar to IRS 13 that is getting tidally disrupted upon the approach towards the SMBH. We investigate this possibility in more detail in this paper. We focus on the area around IRS 1W that contains 42 sources including IRS 1W. The geometrical center of these sources is situated at $6.24 \pm 0.47$ arcsec (right ascension) and $0.25 \pm 0.42$ arcsec (declination) with respect to Sgr A* in 2005.366 epoch within the circular region with a radius of $1.35''$.
 
The paper is structured as follows. In Section~\ref{Observation}, we describe analyzed datasets in the near-infrared (NIR) domain. The identification of stars associated with IRS~1W that constitute an apparent overabundance in comparison with other fields around Sgr~A* is analyzed in Section~\ref{Source identification}. The NIR photometry colour-colour diagram of the sources is presented in Section~\ref{section:color-color diagram}. Proper motions of the sources are analyzed in Section~\ref{proper_motion}, where the significant stellar concentration with the common proper motion in the north-western direction is identified. We discuss potential clustering scenarios (IMBH-bound cluster or a projected disk-like distribution) in Section~\ref{clustering_scenarios}. Finally, we summarize the main results in Section~\ref{section:Conclusion}.


\begin{table}[tbh!]
\vspace*{1em}
\centering
 \begin{tabular}{ rrrr }
 \hline
 \hline   
 Date   & Camera & Filter   &Observation ID\\
   \hline
	2003.451 & S13	  &      $K_s$       	&	713-0078(A)\\
	2003.453 & S13   &  	 $K_s$ 			&	713-0078(A)\\
	2003.456 & S13   &  	 $K_s$ 			&	713-0078(A)	\\
	2003.464 & S13   &  	 $K_s$ 			&	271.B-5019(A)\\
	2003.763 & S13   &  	 $K_s$ 			&	072.B-0285(A)\\
	2004.518 & S13   &  	 $K_s$ 			&	073.B-0775(A)\\
	2004.663 & S13   &  	 $K_s$ 			&	073.B-0775(A)\\
	2004.671 & S13   &  	 $K_s$ 			&	073.B-0775(B)\\
	2004.728 & S13   &  	 $K_s$ 			&	073.B-0085(C)\\
	2005.270 & S13   &  	 $K_s$ 		    &	073.B-0085(I)\\
	2005.366 & S13   &  	 $K_s$ 			&	073.B-0085(D)\\
	2005.371 & S13   &  	 $K_s$ 			&	073.B-0085(D)\\
	2005.374 & S13   &  	 $K_s$ 			&	073.B-0085(D)\\
	2005.467 & S27   &  	 $K_s$ 			&	073.B-0085(F)\\
	2005.569 & S13   &  	 $K_s$ 			&	075.B-0093(C)\\
	2005.574 & S13   &  	 $K_s$ 			&	075.B-0093(C)\\
	2005.577 & S13   &  	 $K_s$ 			&	075.B-0093(C)\\
	2005.585 & S13   &  	 $K_s$ 			&	075.B-0093(C)\\
	2006.585 & S13   &  	 $K_s$ 			&	077.B-0014(D)\\
	2007.253 & S13   &  	 $K_s$ 			&	179.B-0261(C)\\
    2007.256 & S13    &  	 $H  $          & 179.B-0261(A)	 \\
	2009.503 & S13   &  	 $K_s$ 			&	183.B-0100(D)\\
	2009.508 & S13   &  	 $K_s$ 			&	183.B-0100(D)\\
	2010.240 & S13   &  	 $K_s$ 		 	&	183.B-0100(L)\\
	2013.667 & S13   &  	 $K_s$ 			&	091.B-0183(B)\\
	2017.456 & S13   &  	 $K_s$ 			&	598.B-0043(L)\\
	2018.306 & S13   &  	 $K_s$ 			&	598.B-0043(O)\\
	2018.311 & S13   &  	 $K_s$ 			&	0101.B-0052(B)\\
   2008.400 & L27   &        $L'$            &  081.B-0648(A) \\
\hline
\hline
 \end{tabular}
 \caption{Summary of NIR observations in $H$, $K_s$, $L'$ bands including the date, camera, filter, and observation ID. $K_s$-band data is used for proper-motion measurements. Photometric measurements are performed in all of the corresponding bands. The pixel scale of NACO S13 camera is 13 ${\rm mas}/$pixel and that of S27 is 27 ${\rm mas}/$pixel.}
\label{table:observations}
\end{table}

\section{Observations and data reduction}
\label{Observation}

 The Galactic Center observations presented in this paper are based mostly on the infrared datasets, with the addition of the data from the X-ray domain. The infrared observations were carried out using the Nasmyth Adaptive Optics System/Coude NIR Camera (NAOS/CONICA) at the European Southern Observatory (ESO) Very Large Telescope (VLT). The M-type supergiant IRS 7 located at $\sim 5.5''$ north of Sgr~A* \citep{Becklin1975,  Lebofsky1982, Yusef-Zadeh1991} serves as a natural guide star for the adaptive optics system. The X-ray data was obtained using the \textit{Chandra} X-ray observatory. The $K_s$-band data (2.2\,$\mu m$) were taken at Unit 4 (UT4)-YEPUN from $2002$ to $2013$ and from 2014 at UT1-Antu on Paranal, Chile. The $K_s$-band images taken with the S13 camera have a pixel scale of 13\,${\rm mas}/$pixel and there is one data set among our studied data sets in $K_s$-band data taken by the S27 camera which has a pixel scale of 27 ${\rm mas}/$pixel. We use 16 epochs from 2003.451 to 2018.311 (28 observations) to derive the proper motions of a group of stars located at the east side of the bow-shock object IRS 1W in a circular region with the radius of about $1.35''$ (1'' corresponds to 0.04 pc, hence the region has the physical radius of $\sim 54$ milliparsecs). 
 
In this work, for the photometric purposes, we use the $H$-(1.6\,$\mu$m), $K_s$-(2.2\,$\mu$m) and $L'$-(3.8$\mu$m) band data obtained with the NACO@VLT with a pixel scale of 13, 13, and 27 ${\rm mas}/pixel$, respectively, and the X-ray data with the pixel scale of 492\,${\rm mas}/$pixel. 
In Table~\ref{table:observations}, we summarize all the data sets which are a part of our analysis.

The standard data reduction process is applied to all of our infrared data sets that consists of the sky subtraction, flat-fielding, bad and dead pixel correction and finally combining the images for each epoch via a shift-and-add algorithm to obtain the final array. 

The X-ray data span the energy range from 0.5 to 8 keV and from 4 to 8 keV in 2005. The X-ray data reduction follows the procedure as explained in \cite{Mossoux2017}.


\section{Source identification}
\label{Source identification}
\begin{figure}[tbh!]
 \includegraphics[width=1\linewidth]{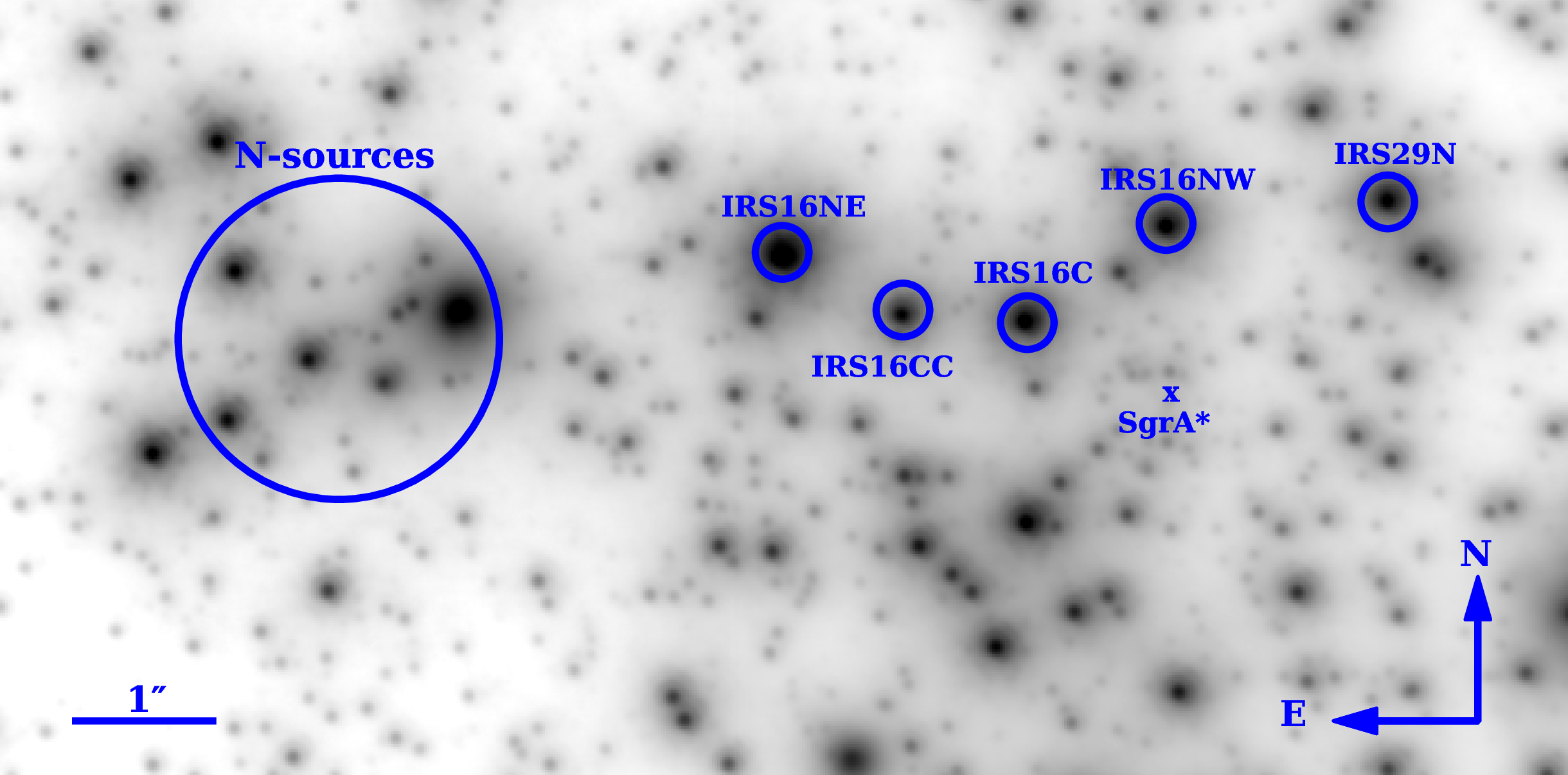}
   \caption{The image obtained in $K_s$-band from 2005.366 epoch. Symbol ``x'' denotes the position of Sgr~A*. The encircled N-sources at the projected distance of $\sim 6.05''$ with respect to Sgr~A* stands for the region of our study.}
\label{fig:full image}
\end{figure}

 We study sources within a circular region located at $\sim 6.05''$ north-east of Sgr~A* with a radius of $\sim 1.35''$ (shown in Fig.~\ref{fig:full image}). In this section, we show that the circular region is not chosen arbitrarily but it rather encompasses the stellar overabundance in the region. By increasing the aperture size, the density of sources drops significantly. In addition, there is a visible gap between the N-sources' region and the IRS16 sources' region.

\begin{figure}[tbh!]
\begin{center}
\includegraphics[width=\linewidth]{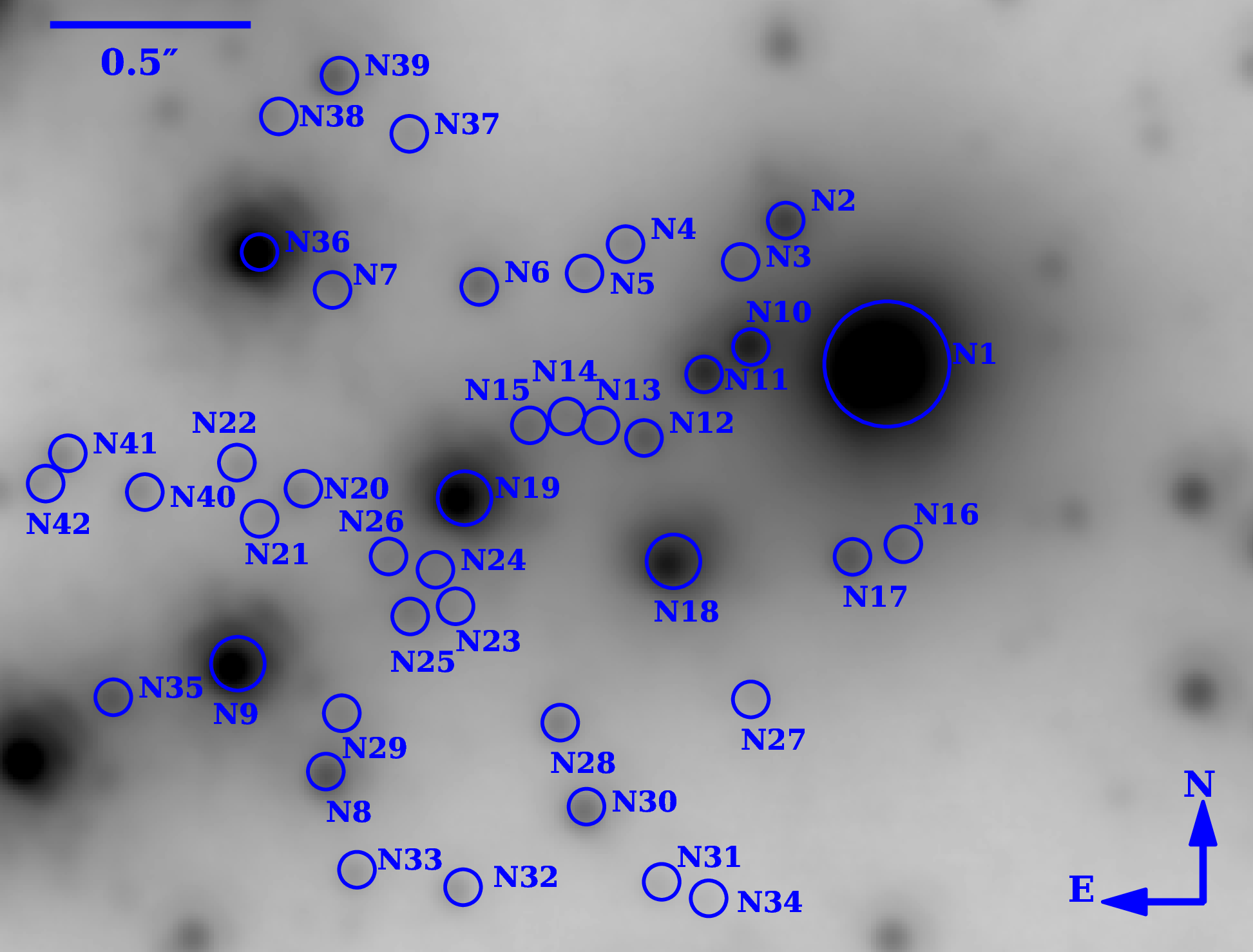}
\caption{The $K_s$-band image from 2005.366 epoch. The sources are labeled from N1/IRS 1W to N42. The figure is a zoom-in of the encircled N-sources' region in Fig.~\ref{fig:full image}, which is located at $\sim 6.05''$ north-east of Sgr~A*.} 
\label{fig:identified sources}
\end{center}
\end{figure}

The brightest star in this region is the aforementioned bow-shock source IRS 1W. We identify 42 sources including IRS 1W in the previously described region. We name the sources starting with the letter N from N1 for IRS 1W to N42. In Fig.~\ref{fig:identified sources}, we show the identified sources in $K_s$-band. We call these sources the N-sources for simplicity. The N-sources are not only identified in the $K_s$-band but also in the $H$- and the $L'$-bands.

We determined the positions of the N-sources using the \texttt{StarFinder} software \citep{Diolaiti2000} in the $K_s$-band high-pass filtered images from 2003.451 to 2018.311 over 16 epochs.

\begin{figure}[tbh!]
\begin{center}
 \includegraphics[width=0.5\textwidth]{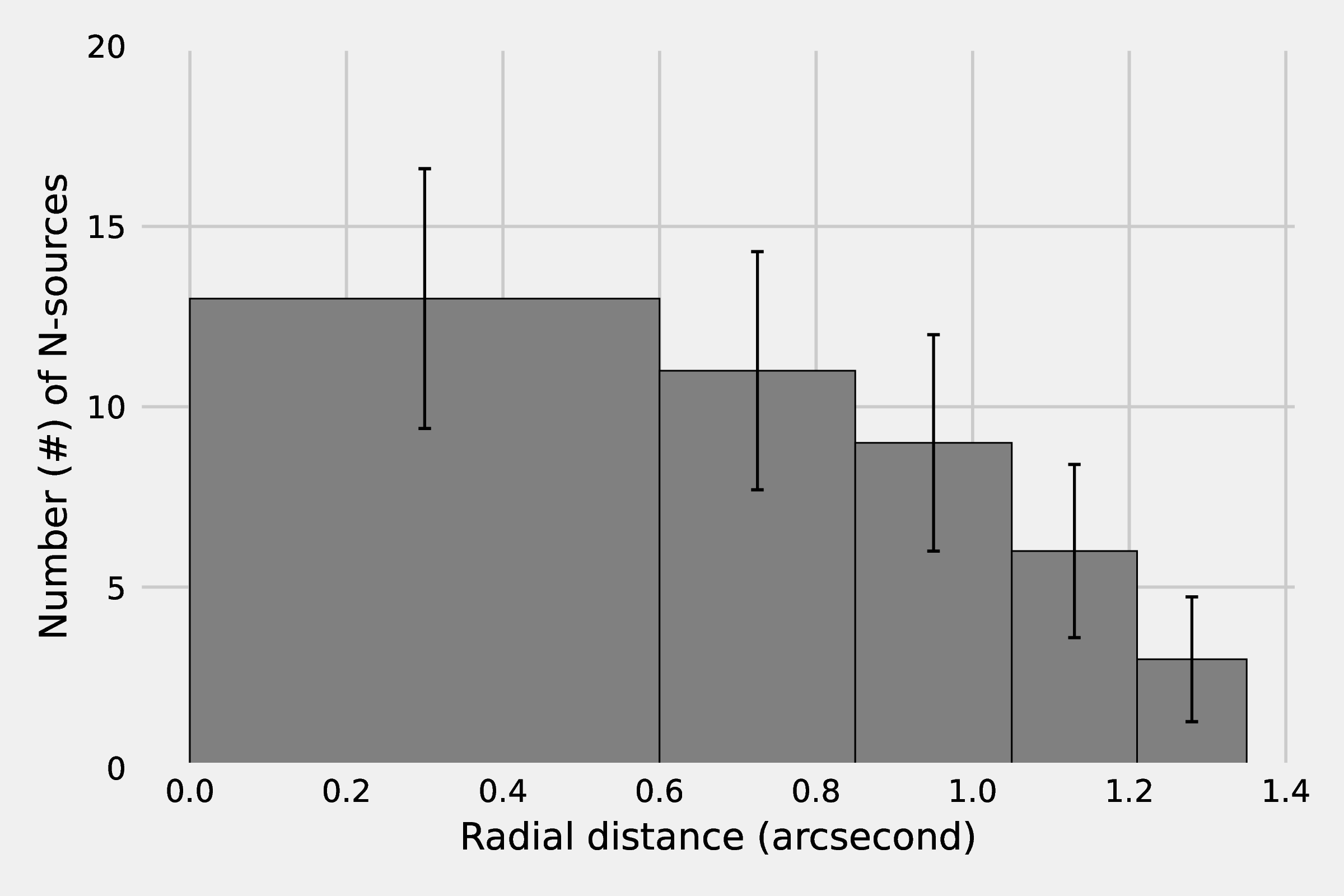}
   \caption{Distribution of $N$-sources as a function of the projected distance with respect to the geometrical center of the association. The $y$-axis denotes the number of sources and the $x$-axis shows the radial distance (in arcseconds) from the geometrical center of N-sources. The width of the bins in the histogram is defined by concentric apertures centered at the geometrical center of N-sources. In addition, each annulus has an equal area to the central disk, with the radius of $0.6''$ for the central disk, and the radial extent of $0.6''$ to $0.8''$ for the first annulus, $0.8''$ to $1.0''$ for the second annulus, $1.0''$ to $1.2''$ for the third annulus, and $1.2''$ to $1.3''$ for the fourth annulus.}
\label{fig:number density of all sources}
\end{center}
\end{figure}

When we investigate the number of the N-sources with respect to the radial distance from the geometrical center of the N-sources (Fig. \ref{fig:number density of all sources}), we find a gradual decrease in the number of the sources with the increasing radial distance. This is an expected trend for at least partially relaxed star cluster with a cusp-like power-law density distribution. 
 Strong variations in source density are unlikely to be caused by
the variable extinction across the field under investigation.
These variations amount to a maximum of $\pm$0.2 in H-K$_s$ and
K$_s$-L following \citet{2010A&A...511A..18S} over the
entire field and even less across the individual regions.

\begin{table}[tbh!]
\centering
 \begin{tabular}{ rr}
 \hline
 \hline
Aperture ID  & number of sources \\
 \hline
N-sources  &   42     \\
 \hline
Region 1   &   15     \\
Region 2   &   13     \\
Region 3   &   13     \\
Region 4   &   16     \\
Region 5   &   16     \\
Region 6   &   17     \\
Region 7   &   14     \\
Region 8   &   14     \\
Region 9   &   12     \\
Region 10   &  15     \\
Region 11   &  16     \\
 \hline
 \hline
 \end{tabular}
 \caption{Number of sources in eleven randomly chosen regions with a comparable distance from Sgr~A* and a radius to the N-source association region. Each region is located at a distance of about $6.05''$ from Sgr~A* and its radius is $\sim 1.35''$. The mean number of sources in these 11 regions is $14.6 \pm 1.5$}
    \label{table:11 regions}
\end{table}

To test the cluster hypothesis for the N-sources, it is crucial to compare their number density with random areas distributed isotropically at a comparable distance from Sgr~A*. A stable cluster or a stellar association in the Galactic Center must necessarily be characterized by an overdensity with respect to other comparable regions. To this goal, we select eleven random circular regions with two conditions. First, the center of each region should have the same distance from Sgr~A* as the distance of the N-sources' region. Second, each of these circular regions should have the same radius as the chosen aperture representing the N-source region. These comparison regions are labeled from ``Region 1'' to ``Region 11'' as shown in the Appendix (Fig.~\ref{fig:11 regions}). The number of detected stars in $K_s$-band in each region is listed in Table~\ref{table:11 regions}. The mean number of sources in our eleven random samples is $14.6 \pm 1.5$. The number of sources in our 11 regions shows that the N-source region is characterized by a significant stellar overdensity, which leads us to speculate that the N-sources or their subgroup form a bound stellar association. 

We cross-checked the identified N sources with the previous proper-motion and radial-velocity studies \citep{2009A&A...502...91S,2016ApJ...821...44F,2022ApJ...932L...6V} that include stars in the N source region. There are 13 corresponding sources in \citet{2009A&A...502...91S}, out of which 8 move northward, 8 common sources in \citet{2016ApJ...821...44F}, out of which 5 share north-ward direction, and 5 common sources in \citet{2022ApJ...932L...6V} that all move north-ward. Hence, for the majority of the common sources, there is consistency in the proper motion. Differences arise mostly for fainter sources (ten times fainter than the brightest object), which are affected by larger uncertainties in the identification in the crowded field.  In \citet{2009A&A...502...91S} the images were subjected to a PSF fitting
routine in order to extract the point source positions. In \citet{2016ApJ...821...44F} the positions were extracted from Lucy deconvolved
images \citep[see also][and references therein]{2022ApJ...932L...6V}.
The high pass filtering method applied in this work is a linear
process in which all image scales and distances are preserved.
No PSF fitting is involved, yet the bulk of the PSF is removed and the high-resolution image cores of the stars are preserved.
It is free from potential deconvolution artifacts that may
occur especially for fainter sources in particular
if the number of iterations is kept too small.
It is also free from potential uncertainties that may occur if
the PSF is fitted in particular to faint sources in a crowded field. In addition, the temporal baseline in this work is 15 years, while it is more limited in the previous works \citep[by a factor of 2.5 longer than the baseline in][]{2009A&A...502...91S}. 

\section{Colour-colour Diagram}
\label{section:color-color diagram}

\begin{figure}[tbh!]
\begin{center}
\includegraphics[width=1\linewidth]{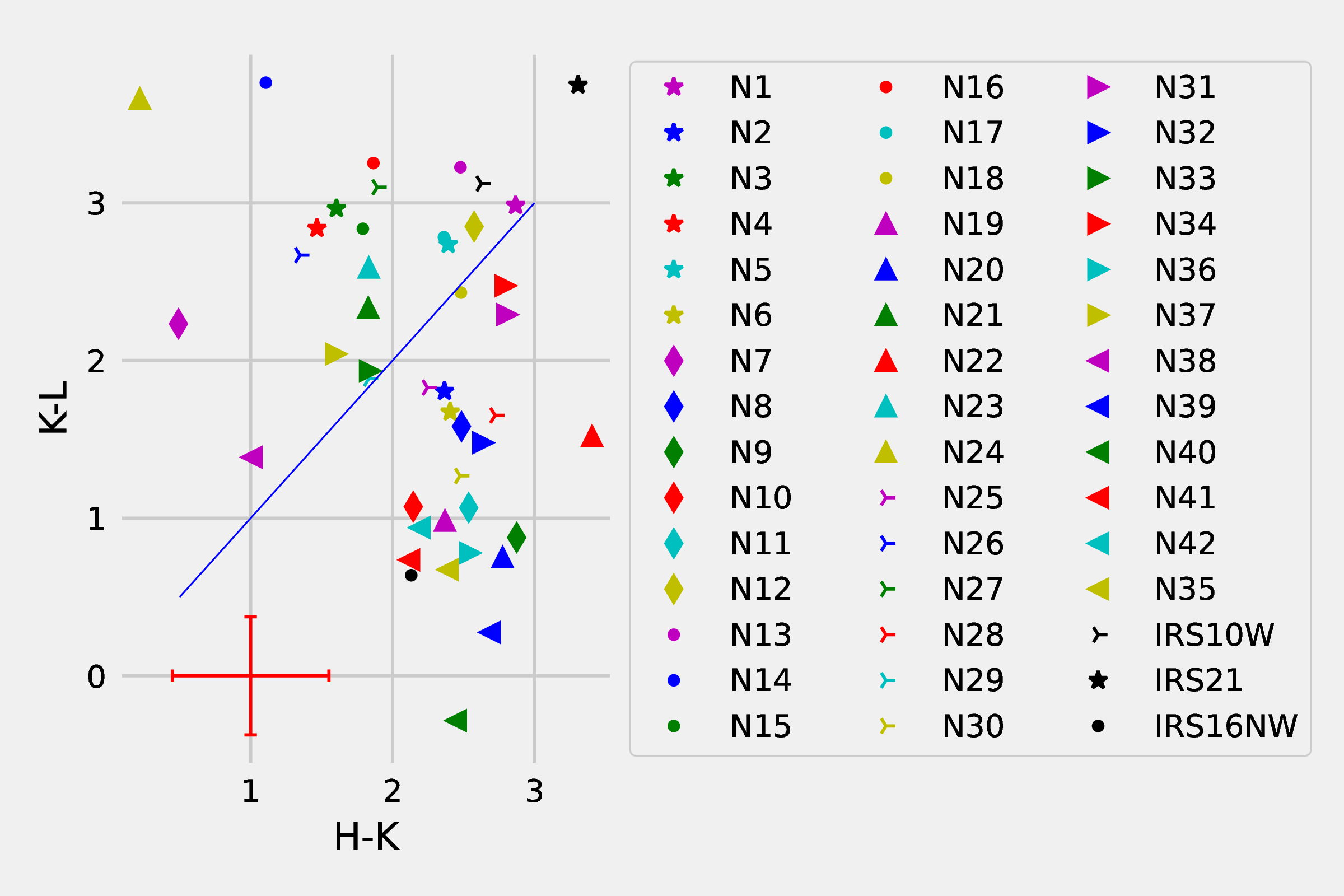}
\caption{HKL two-colour diagram. The solid blue line denotes the colors of a one-component black body at different temperatures. The red cross represents the uncertainty of the values. The red cross denotes the mean photometric uncertainties. }
\label{fig:color-color diagram}
\end{center}
\end{figure}

The flux densities of N-sources in $H$, $K_s$ and $L'$-bands are obtained using aperture photometry with an aperture radius of 4, 4 and 3 pixels in $H$, $K_s$ and $L'$-bands, respectively. Our reference stars for photometry are IRS 10W, IRS21 and IRS16NW and their dereddened magnitudes are taken from \citet{Blum1996}. In order to optimize the background subtraction for each source and reference stars, we use two apertures around the source with the same size as photometry apertures. The flux densities for Vega in $H$, $K_s$- and $L'$-bands are 1050, 667 and 248 ${\rm Jy}$, respectively. All N-sources are detected in $H$, $K_s$ and $L'$-bands. We list their corresponding dereddened magnitudes in Table~\ref{table:Magnitudes} and the derived colour indices $H-K_s$ and $K_s-L'$ in Table~\ref{table:Color-color}. The results are presented in the colour-colour diagram, see Fig.~\ref{fig:color-color diagram}. 

It cannot be excluded that several (e.g.\ four) of the bluest stars could be
foreground stars.
In the literature, the cut to identify foreground stars is often
set around 1.3 mag -- 1.8 mag \citep{2010A&A...511A..18S,2016ApJ...821...44F}.
However, the foreground extinction is also variable and on
spatial scales of less than 1 arcsecond typical angular resolution of the extinction maps it cannot be excluded
that these blue stars (independent of their brightness) are local
cluster stars stars with exceptionally low foreground extinction.
However, for the faint sources  presented here the stellar photometry
has large uncertainties and the bluest stars
could still be consistent with being in the Galactic center - with the possible exception of N24 with a rather blue H-K value. Given these facts, we assume that the region is free of genuine foreground stars that are not part of the 
Galactic Center nuclear stellar cluster.

An interpretation of the colour-colour diagram shown in Fig.~\ref{fig:color-color diagram}
is possible in the framework of the corresponding diagrams
shown in Fig.~16 of \citet{Eckart2013}.
Within the uncertainties, the sources that are located to the right
of the single-component black-body line are objects reddened by the extinction towards
the Galactic Center region. Objects that are located above the
black-body line experience additional local extinction due to
dust locally present in or in front of the IRS 1W region.

\section{Proper motion}
\label{proper_motion}

For the first time, we derive and present proper motions of N-sources. The prevailing angle of the proper motion is northward, which suggests that N-sources could be a co-moving group of stars.

In order to study the proper motions of candidate stars, we identified the N-sources in $K_s$-band images from 2003 to 2018. With the aid of the S2 star position in each epoch, the position of each source was transformed into the common coordinate system with the central position of Sgr~A*.
As an example, we show the best-fitting proper motions based on the derived positions as a function of time in right-ascension (R.A.) and declination (Dec.) directions in Fig. \ref{plot:proper motion of w2} for the star N2. 

\par We adopt the distance of $8$ kpc to Sgr~A* \citep{Shapley1928, Reid1993, Eisenhauer2003, Gravity-distance2019} to transform the proper motions to velocities. The inferred velocities are listed in Table~\ref{table:proper motion wsources km/s} alongside the R.A. and Dec. coordinates with respect to Sgr~A* for all the sources. IRS 1W, IRS 16C, IRS 16CC, IRS 16NE, and IRS 16NW are adopted as reference stars (Table~\ref{table:velocities of reference stars}), which show an agreement in terms of the inferred velocities with the presented values in \citet{Paumard2006} within uncertainties.

\begin{table*}
\begin{center}
 \begin{tabular}{crrrrrr}
 \hline
 \hline
Name  & $\Delta\alpha$ ($arcsec$) & $\Delta \delta$ ($arcsec$)  & $V_{\rm R.A.}$ \textit{(km/s)} & $V_{\rm Dec.}$ \textit{(km/s)} & $V$ \textit{(km/s)} & Angle \textit{(degree)}\\
\hline
N1	&	5.15	$\pm$	0.01	&	0.58	$\pm$	0.01	&	-89.11	$\pm$	16.31	&	349.24	$\pm$	43.99	&	360.43	$\pm$	42.81	&	-14.31	$\pm$	8.56	\\
N2	&	5.39	$\pm$	0.01	&	0.95	$\pm$	0.01	&	-72.43	$\pm$	10.62	&	336.35	$\pm$	36.78	&	344.06	$\pm$	36.03	&	-12.15	$\pm$	6.09	\\
N3	&	5.51	$\pm$	0.01	&	0.84	$\pm$	0.01	&	-48.16	$\pm$	28.06	&	111.11	$\pm$	31.85	&	121.10	$\pm$	31.28	&	-23.43	$\pm$	35.88	\\
N4	&	5.80	$\pm$	0.01	&	0.89	$\pm$	0.01	&	-86.84	$\pm$	15.93	&	304.88	$\pm$	27.68	&	317.01	$\pm$	26.98	&	-15.90	$\pm$	8.30	\\
N5	&	5.91	$\pm$	0.01	&	0.82	$\pm$	0.01	&	-115.28	$\pm$	13.65	&	271.89	$\pm$	32.23	&	295.32	$\pm$	30.15	&	-22.98	$\pm$	9.80	\\
N6	&	6.18	$\pm$	0.01	&	0.79	$\pm$	0.01	&	-180.12	$\pm$	15.55	&	479.31	$\pm$	29.96	&	512.04	$\pm$	28.57	&	-20.60	$\pm$	5.62	\\
N7	&	6.53	$\pm$	0.01	&	0.77	$\pm$	0.01	&	-63.33	$\pm$	34.51	&	99.35	$\pm$	45.50	&	117.82	$\pm$	42.62	&	-32.52	$\pm$	49.92	\\
N8	&	6.57	$\pm$	0.01	&	-0.49	$\pm$	0.01	&	97.83	$\pm$	12.51	&	299.19	$\pm$	43.23	&	314.78	$\pm$	41.27	&	18.11	$\pm$	0.57	\\
N9	&	6.82	$\pm$	0.01	&	-0.20	$\pm$	0.01	&	-89.11	$\pm$	16.31	&	349.62	$\pm$	43.99	&	360.80	$\pm$	42.82	&	-14.3	$\pm$	8.55	\\
N10	&	5.48	$\pm$	0.01	&	0.63	$\pm$	0.01	&	-121.34	$\pm$	9.86	&	521.78	$\pm$	25.41	&	535.70	$\pm$	24.85	&	-13.09	$\pm$	3.29	\\
N11	&	5.59	$\pm$	0.01	&	0.56	$\pm$	0.01	&	-96.70	$\pm$	12.51	&	185.05	$\pm$	25.79	&	208.79	$\pm$	23.58	&	-27.59	$\pm$	12.67	\\
N12	&	5.75	$\pm$	0.01	&	0.39	$\pm$	0.01	&	-269.23	$\pm$	14.79	&	203.25	$\pm$	28.06	&	337.34	$\pm$	20.62	&	-52.95	$\pm$	10.6	\\
N13	&	5.89	$\pm$	0.01	&	0.43	$\pm$	0.01	&	61.43	$\pm$	15.55	&	401.95	$\pm$	27.68	&	406.62	$\pm$	27.46	&	8.69	$\pm$	3.17	\\
N14	&	5.96	$\pm$	0.01	&	0.44	$\pm$	0.01	&	129.31	$\pm$	17.82	&	497.89	$\pm$	27.68	&	514.41	$\pm$	27.16	&	14.56	$\pm$	2.30	\\
N15	&	6.07	$\pm$	0.01	&	0.41	$\pm$	0.01	&	-48.16	$\pm$	15.93	&	200.22	$\pm$	32.23	&	205.93	$\pm$	31.56	&	-13.52	$\pm$	12.99	\\
N16	&	5.10	$\pm$	0.01	&	0.12	$\pm$	0.01	&	106.56	$\pm$	21.24	&	236.24	$\pm$	31.47	&	259.16	$\pm$	29.99	&	24.28	$\pm$	2.90	\\
N17	&	5.22	$\pm$	0.01	&	0.08	$\pm$	0.01	&	186.19	$\pm$	16.31	&	324.60	$\pm$	22.37	&	374.21	$\pm$	21.03	&	29.84	$\pm$	0.93	\\
N18	&	5.69	$\pm$	0.01	&	0.06	$\pm$	0.01	&	-32.61	$\pm$	17.44	&	580.93	$\pm$	37.92	&	581.84	$\pm$	37.87	&	-3.21	$\pm$	3.87	\\
N19	&	6.24	$\pm$	0.01	&	0.23	$\pm$	0.01	&	-98.97	$\pm$	9.48	&	308.67	$\pm$	42.85	&	324.15	$\pm$	40.91	&	-17.78	$\pm$	7.91	\\
N20	&	6.66	$\pm$	0.01	&	0.24	$\pm$	0.01	&	195.67	$\pm$	14.79	&	304.12	$\pm$	32.23	&	361.63	$\pm$	28.26	&	32.76	$\pm$	1.60	\\
N21	&	6.73	$\pm$	0.01	&	0.18	$\pm$	0.01	&	243.07	$\pm$	16.68	&	295.78	$\pm$	40.95	&	382.84	$\pm$	33.36	&	39.41	$\pm$	3.97	\\
N22	&	6.80	$\pm$	0.01	&	0.31	$\pm$	0.01	&	-59.53	$\pm$	10.24	&	286.68	$\pm$	47.02	&	292.80	$\pm$	46.08	&	-11.73	$\pm$	7.83	\\
N23	&	6.25	$\pm$	0.01	&	-0.06	$\pm$	0.01	&	-91.39	$\pm$	20.86	&	306.77	$\pm$	38.68	&	320.09	$\pm$	37.55	&	-16.59	$\pm$	11.18	\\
N24	&	6.31	$\pm$	0.01	&	0.04	$\pm$	0.01	&	-73.56	$\pm$	28.44	&	447.08	$\pm$	36.40	&	453.09	$\pm$	36.21	&	-9.34	$\pm$	8.62	\\
N25	&	6.36	$\pm$	0.01	&	-0.06	$\pm$	0.01	&	-38.68	$\pm$	16.31	&	273.40	$\pm$	31.85	&	276.12	$\pm$	31.62	&	-8.05	$\pm$	8.64	\\
N26	&	6.43	$\pm$	0.01	&	0.07	$\pm$	0.01	&	-131.96	$\pm$	30.34	&	428.50	$\pm$	47.02	&	448.36	$\pm$	45.82	&	-17.12	$\pm$	10.99	\\
N27	&	5.47	$\pm$	0.01	&	-0.26	$\pm$	0.01	&	-81.15	$\pm$	17.06	&	211.21	$\pm$	28.44	&	226.26	$\pm$	27.24	&	-21.02	$\pm$	13.28	\\
N28	&	5.98	$\pm$	0.01	&	-0.35	$\pm$	0.01	&	53.09	$\pm$	14.41	&	503.96	$\pm$	29.58	&	506.75	$\pm$	29.46	&	6.01	$\pm$	2.55	\\
N29	&	6.54	$\pm$	0.01	&	-0.33	$\pm$	0.01	&	74.70	$\pm$	16.68	&	150.54	$\pm$	37.54	&	168.05	$\pm$	34.44	&	26.39	$\pm$	1.27	\\
N30	&	5.91	$\pm$	0.01	&	-0.58	$\pm$	0.01	&	-95.94	$\pm$	15.93	&	190.74	$\pm$	28.44	&	213.51	$\pm$	26.40	&	-26.70	$\pm$	14.53	\\
N31	&	5.72	$\pm$	0.01	&	-0.77	$\pm$	0.01	&	-42.47	$\pm$	19.34	&	301.46	$\pm$	30.72	&	304.44	$\pm$	30.54	&	-8.02	$\pm$	8.88	\\
N32	&	6.24	$\pm$	0.01	&	-0.78	$\pm$	0.01	&	50.43	$\pm$	19.34	&	284.78	$\pm$	30.72	&	289.21	$\pm$	30.44	&	10.04	$\pm$	5.49	\\
N33	&	6.51	$\pm$	0.01	&	-0.74	$\pm$	0.01	&	117.17	$\pm$	28.06	&	720.86	$\pm$	37.54	&	730.32	$\pm$	37.33	&	9.23	$\pm$	3.41	\\
N34	&	5.61	$\pm$	0.01	&	-0.80	$\pm$	0.01	&	156.61	$\pm$	19.34	&	396.64	$\pm$	40.57	&	426.44	$\pm$	38.40	&	21.55	$\pm$	0.84	\\
N35	&	7.14	$\pm$	0.01	&	-0.29	$\pm$	0.01	&	-77.36	$\pm$	22.75	&	289.33	$\pm$	50.81	&	299.49	$\pm$	49.44	&	-14.97	$\pm$	13.65	\\
N36	&	6.77	$\pm$	0.01	&	0.87	$\pm$	0.01	&	-21.24	$\pm$	18.96	&	353.41	$\pm$	45.5	&	354.05	$\pm$	45.43	&	-3.44	$\pm$	7.11	\\
N37	&	6.37	$\pm$	0.01	&	1.17	$\pm$	0.01	&	-257.10	$\pm$	34.13	&	197.18	$\pm$	56.12	&	324.01	$\pm$	43.59	&	-52.51	$\pm$	22.80	\\
N38	&	6.73	$\pm$	0.01	&	1.22	$\pm$	0.01	&	144.10	$\pm$	22.75	&	319.29	$\pm$	72.05	&	350.30	$\pm$	66.34	&	24.29	$\pm$	3.05	\\
N39	&	6.56	$\pm$	0.01	&	1.33	$\pm$	0.01	&	-28.06	$\pm$	18.58	&	519.88	$\pm$	58.78	&	520.64	$\pm$	58.70	&	-3.09	$\pm$	4.84	\\
N40	&	7.06	$\pm$	0.01	&	0.25	$\pm$	0.01	&	-38.68	$\pm$	10.62	&	487.65	$\pm$	37.92	&	489.18	$\pm$	37.81	&	-4.54	$\pm$	3.20	\\
N41	&	7.27	$\pm$	0.01	&	0.33	$\pm$	0.01	&	122.10	$\pm$	17.06	&	155.09	$\pm$	35.27	&	197.39	$\pm$	29.65	&	38.21	$\pm$	5.07	\\
N42&	7.32	$\pm$	0.01	&	0.27	$\pm$	0.01	&	-200.60	$\pm$	51.57	&	477.79	$\pm$	118.31	&	518.19	$\pm$	110.9	&	-22.78	$\pm$	21.01	\\

  \hline
  \hline
 \end{tabular}
 \caption{Velocities of N-sources over the 16 years of monitoring. N1 represents the IRS 1W bow-shock source. The positions of N-sources are given with respect to the position of Sgr~A* in 2005 epoch.}
 \label{table:proper motion wsources km/s}
\end{center} 
\end{table*}

\begin{table}[tbh!]
\centering
 \begin{tabular}{ c rr}
 \hline
 \hline
Name  &  $V_{\rm R.A.}$ \textit{(km/s)} & $V_{\rm Dec.}$ \textit{(km/s)} \\
 \hline
IRS 1W& -89.11 $\pm$ 16.31  &    349.24 $\pm$	43.99 \\     
IRS16C&  -348.48$\pm$ 5.31&	 288.19 $\pm$	8.34      \\ 
IRS16CC& -95.56 $\pm$ 5.31&	    266.96 $\pm$	14.03  \\    
IRS16NE& 84.18 $\pm$ 10.62&	    -283.64	$\pm$ 21.61  \\     
IRS16NW& 195.29 $\pm$ 6.83&	    19.30 $\pm$	4.17   \\   
  \hline
  \hline
 \end{tabular}
 \caption{R.A. and Dec. velocities of the reference stars which are within uncertainties in agreement with \cite{Paumard2006}.}
 \label{table:velocities of reference stars}
\end{table}

\begin{figure}[tbh!]
\begin{center}
\includegraphics[width=\linewidth]{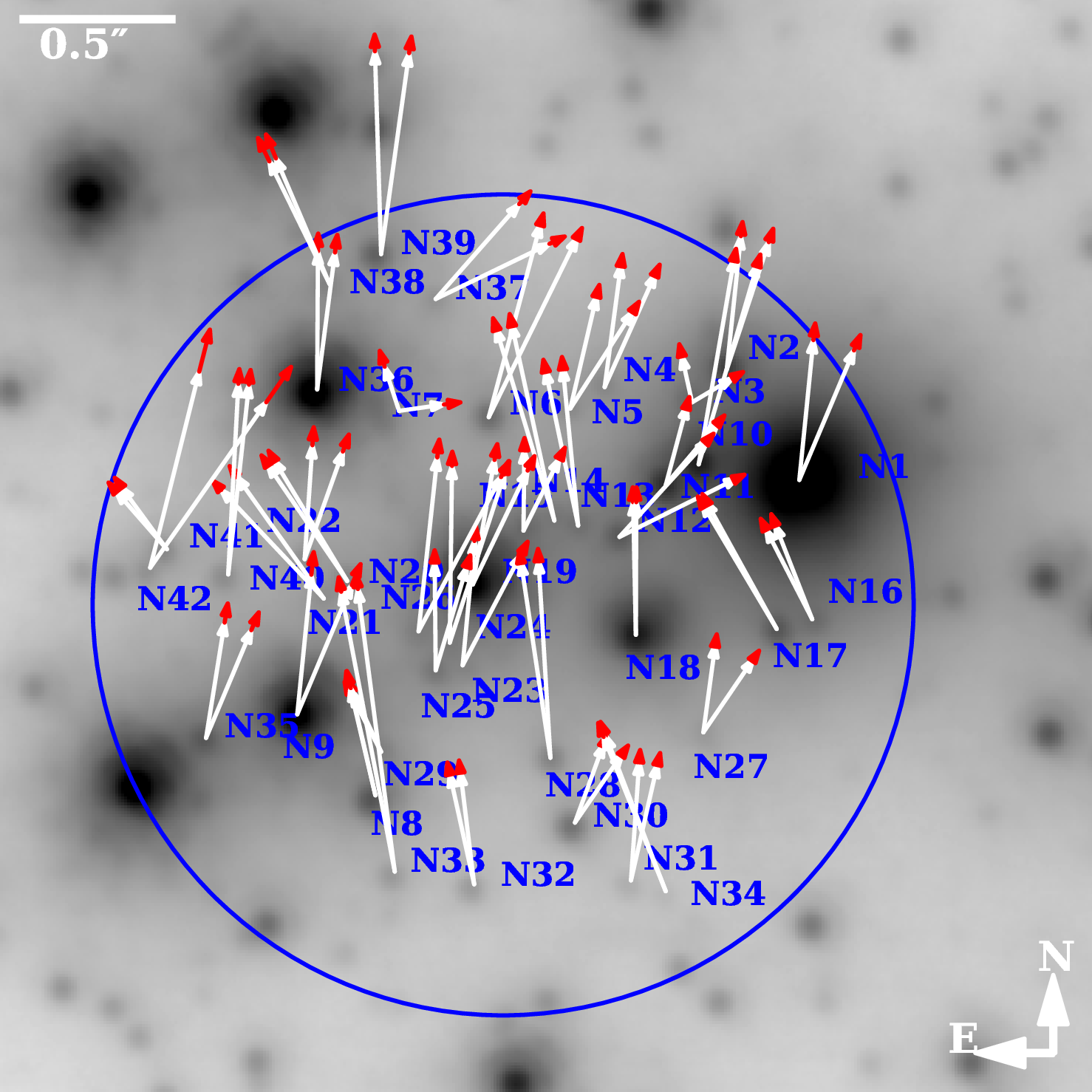}
\caption{The $K_{\rm s}$-band image from 2005.366 epoch with superimposed proper motions for each $N$ source. Four arrows are shown for each source. The white arrows demonstrate the velocity vectors of each source including the uncertainty of the velocity angle, whereas the red arrows indicate the uncertainty in the velocity magnitude. The size of the arrows is proportional to the velocity magnitude. The proper motion vectors are based on Table~\ref{table:proper motion wsources km/s} and the arrows are one order of magnitude smaller than proper motions. In total, 42 sources move North-ward (N-sources).}
\label{fig:N-sources proper motion}   
\end{center}
\end{figure}\textbf{}

 \begin{figure}[tbh!]
 \begin{center}
 \includegraphics[width=0.5\textwidth]{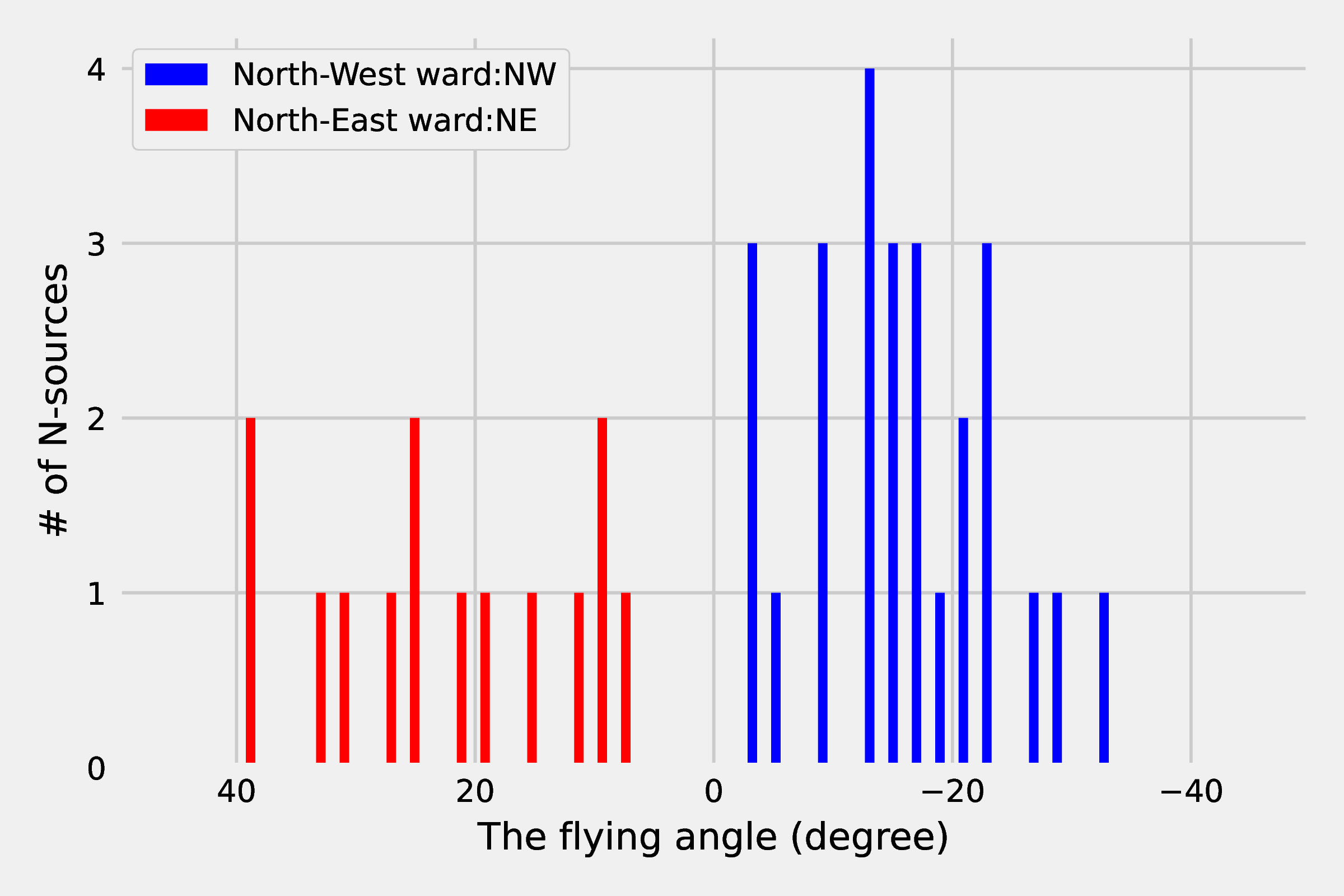}
   \caption{Distribution of flying angles of $N$-sources. The zero angle stands for a Northward flying angle. The negative angles indicate the North-West-ward (NW) flying angles up to $-35$ degrees, whereas the positive angles represent the North-East-ward (NE) flying angles up to about $40$ degrees. There is an indication that the NW-ward sources are members of a denser association of stars, while the rest of the sources might be the foreground/background sources that are still part of the nuclear star cluster.}
\label{fig:histogram_angel}
\end{center}
\end{figure}

\par While the brighter sources in the overall fields show good agreement with proper motion measurements taken from the literature (see Table~\ref{table:velocities of reference stars}), the predominant portion of the stars given in Table~\ref{table:proper motion wsources km/s} have no reported data in the literature 
\citep[see otherwise plentiful sets given by][]{2009ApJ...697.1741B,2014ApJ...783..131Y,2015A&A...584A...2F,2022ApJ...932L...6V}.  Some of them have high velocities and may either be foreground stars or only loosely bound to the overall cluster. Here, data over a longer baseline in time and/or spectroscopy data with high sensitivity and angular resolution are required.

The velocity dispersion in R.A., corrected for the mean error of $19.0\,{\rm km\,s^{-1}}$, is $\sigma_{\rm RA}=58.6^{+2.3}_{-1.4}\,{\rm km\,s^{-1}}$, and in Dec., corrected for the mean error of $39.3\,{\rm km\,s^{-1}}$, is $\sigma_{\rm Dec}=127.3^{+3.4}_{-1.5}\,{\rm km\,s^{-1}}$. The fact that the mean dispersion in Dec. direction is larger by a factor of two may point at an increased source density in that direction. This would support a disk solution as discussed later. Such a source density enhancement could manifest itself in background variations and therefore difficulties in source positioning and determination of the proper motions. However, as we show later for the NW sources, when we remove a few sources exceeding the escape velocity (these are fainter sources), the dispersion in declination approaches the one in right ascension. The total on-sky velocity dispersion is $\sigma_{\star}=128.0^{+4.1}_{-2.5}\,{\rm km\,s^{-1}}$. When we remove four sources classified as late-type stars by \citet{2016ApJ...821...44F} (N6, N9, N28, and N30 with the determined line-of-sight velocities), the velocity dispersion values in R.A., Dec., and the total on-sky velocity dispersion are not affected significantly and are within the values for the whole N cluster ($\sigma_{\rm RA}=59.9\,{\rm km\,s^{-1}}$, $\sigma_{\rm Dec}=127.0\,{\rm km\,s^{-1}}$, and $\sigma_{\star}=122.0\,{\rm km\,s^{-1}}$).

In Fig. \ref{fig:N-sources proper motion}, we show the proper motions of the N-sources superposed on a $K_{\rm s}$-band (2.2\,$\mu$m) image. 
In general, these sources move towards either north-east or north-west. In Fig. \ref{fig:histogram_angel}, we plot the distribution of the N-sources with respect to their flying angle, which indicates two distinct populations of stars. The North-West (NW) flying sources are depicted in blue, whereas the North-East flying ones in red. The Gaussian-like distribution of the number of the North-West flying sources with respect to the flying angle supports the clustering model of these sources. On the other hand, North-East (NE) flying sources exhibit a flat distribution of the number of sources with respect to the flying angle. In summary, 28 sources out of the total 42 identified N-sources move NW-ward (N1, 2, 3, 4, 5, 6, 7, 9, 10, 11, 12, 15, 18, 19, 22, 23, 24, 25, 26, 27, 30, 31, 35, 36, 37, 39, 40, 42), while the rest (14 sources) move NE-ward (N8, 13, 14, 16, 17, 20, 21, 28, 29, 32, 33, 34, 38, 41). The number of NE flying sources is consistent with the mean number in random test regions, see Table~\ref{table:11 regions}, which suggests that they mostly belong to the background/foreground nuclear star cluster population. 
The possible expectation of a broader range of angles for the random NE flying sources that would not belong to the cluster is hampered by the fact that these may still be influenced by a more coherent motion in the disk direction.
On the other hand, 28 NW-ward flying sources constitute a significant overabundance that we investigate in more detail in the following Section.

\begin{figure}[tbh!]
\begin{center}
 \includegraphics[width=0.5\textwidth]{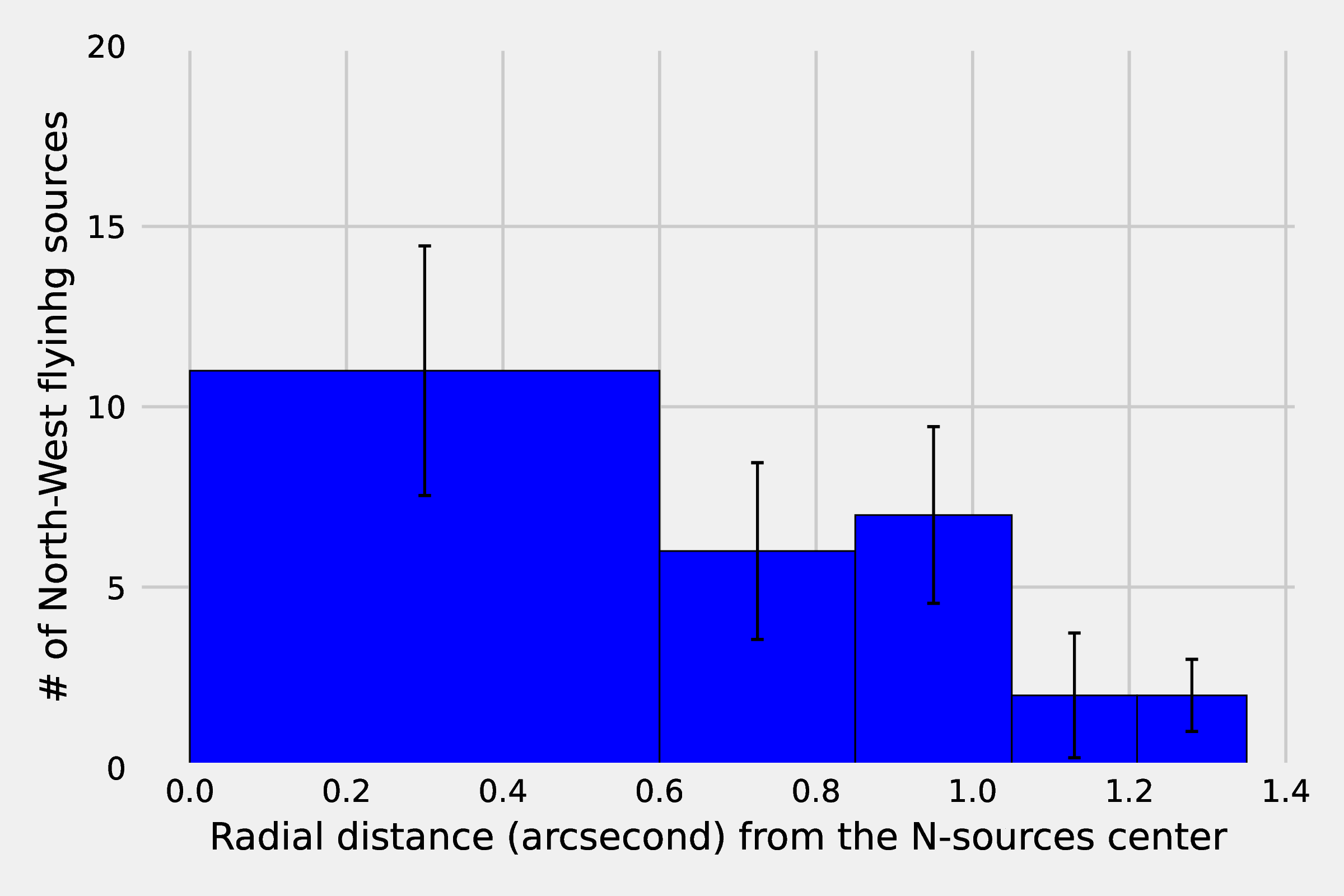}
   \caption{The spatial distribution of the NW-ward flying sources. The zero point of the $x$-axis is at the geometrical center of the N-sources. The spatial distribution is Gaussian-like. }
\label{fig:NW nubmer density_initial center} 
\end{center}
\end{figure}

\begin{figure}[tbh!]
\begin{center}
 \includegraphics[width=0.5\textwidth]{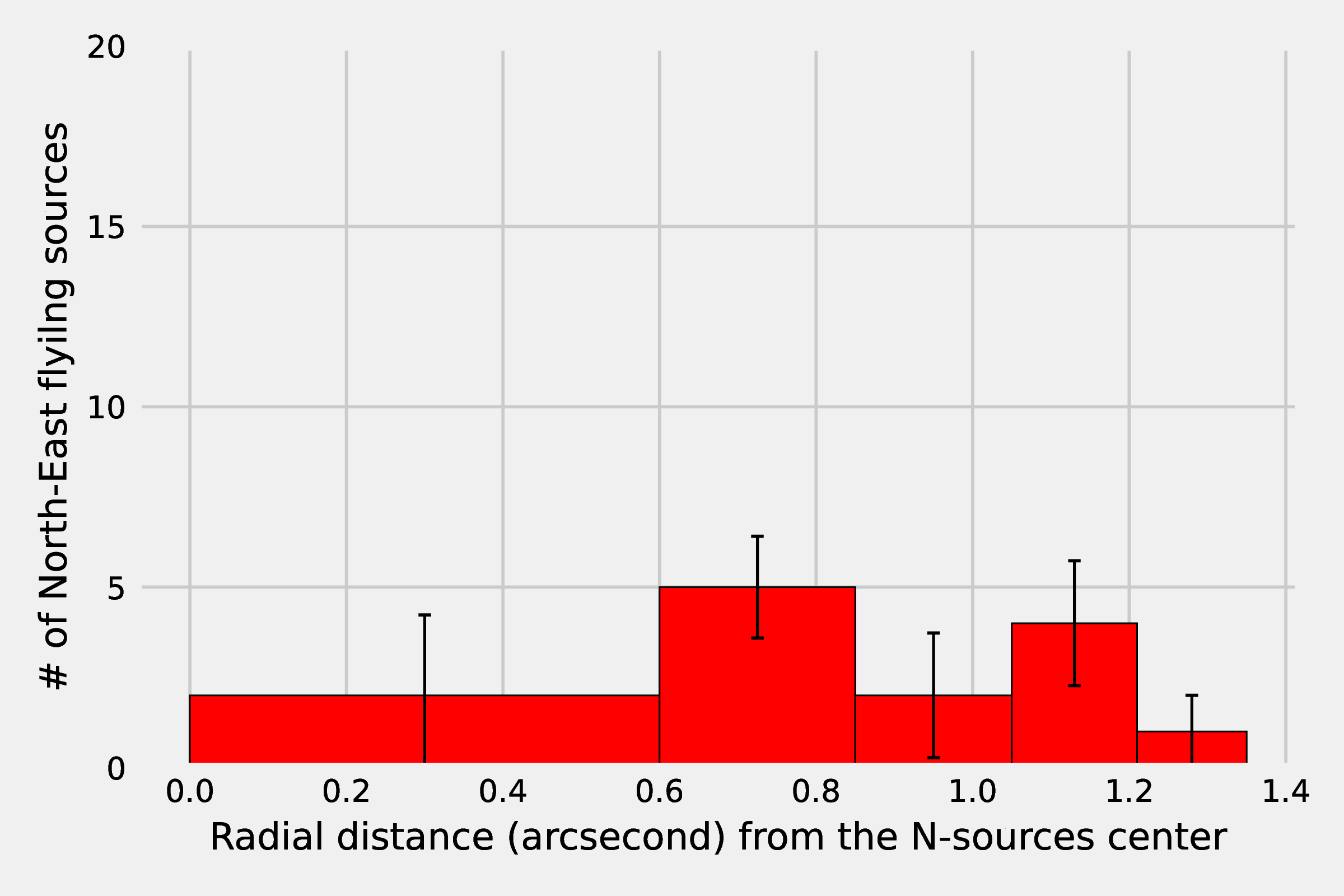}
   \caption{The spatial distribution of the NE-ward flying sources. The zero point of the $x$-axis is at the geometrical center of the N-sources. The spatial distribution is flat within the uncertainties.}
\label{fig:NE nubmer density_initial center}
\end{center}
\end{figure}

\section{Clustering scenarios}
\label{clustering_scenarios}


In the previous section, with the support of Fig.~\ref{fig:histogram_angel}, we showed that the N-sources can be divided into two kinematically distinct categories: more abundant NW-ward flying sources and the NE-ward flying sources consisting most likely of foreground/background sources within the nuclear stellar cluster but not belonging to a possible small cluster associated with the region. In Figs.~\ref{fig:NW nubmer density_initial center} and Fig. \ref{fig:NE nubmer density_initial center}, we show the number of sources as a function of radial distance for NW-ward and NE-ward flying sources, respectively. In these two figures, we kept the geometrical center of the N-sources as the center of each population (see Table~\ref{table:Geometric centers}). In Fig.~\ref{fig:NW nubmer density_initial center}, we demonstrate that the spatial distribution of the NW-ward flying sources is Gaussian-like, see also the fit in the Appendix (Figure~\ref{fig:Gaussian fit}), while the spatial distribution of NE-ward flying sources is flat within uncertainties (see Fig.~\ref{fig:NE nubmer density_initial center}). Thus, based on the flying-angle distribution as well as on the spatial distribution of the NW-ward sources, we can speculate that these sources exhibit the characteristics of a stellar cluster or an association. Therefore, we discuss two plausible scenarios which can result in a bound system or an apparent overdensity. As the first scenario, we propose the existence of an intermediate-mass black hole (IMBH) (see Subsect.~\ref{subsection_IMBH}). In the second scenario, we explain the apparent overdensity on the sky as a result of an inclined disk-like distribution of stars (see Subsection~\ref{section:Disk projection}). 

\begin{figure}[tbh!]
\begin{center}
\includegraphics[width=1\linewidth]{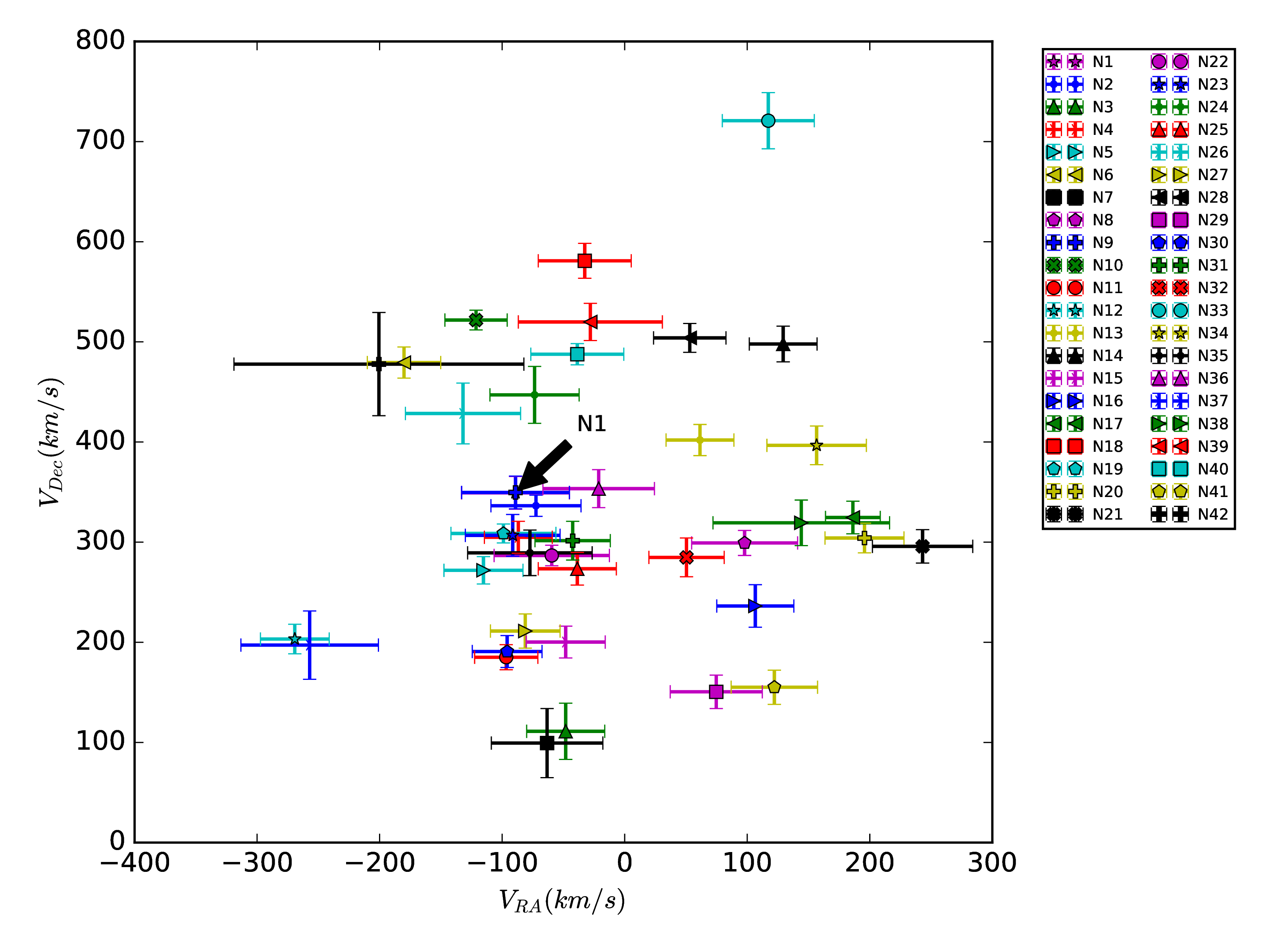}
\caption{In this vector-point diagram we denote the spatial distribution of velocities. The position of N1/ IRS 1W is depicted with a black arrow. }
\label{fig:Vra_Vdeclination diagram}
\end{center}
\end{figure}

The clustering hypothesis is also supported by the spatial distribution of velocities in a vector-point diagram as shown in Fig.~\ref{fig:Vra_Vdeclination diagram}. Here, the strongest clustering occurs around N1, i.e. IRS1W, for sources moving in NW direction (negative RA proper motions). The more extended distribution of sources moving predominantly NE constitute the background source or the source in the GC stellar disk.

\subsection{Putative IMBH}
\label{subsection_IMBH}

In the first scenario, we obtain the mass estimate of a putative IMBH for the association of NW-ward flying sources. Our co-moving group of stars has a random velocity dispersion distribution. Therefore, the center of mass of the NW-ward flying association could be associated with a massive object that would prevent this group from tidal disruption. We assume that the hypothetical IMBH is located close to the geometrical center of our group of stars. In order to estimate the characteristic size of the cluster, we obtain the geometrical center of the NW-ward sources and consider the standard deviation of the source distribution as the characteristic radius of the cluster, which leads to the estimate of $R_{\rm c}=0.57'' \pm 0.10''$\footnote{ In order to fit the distribution in Fig.~\ref{fig:NW nubmer density_initial center} using the Gaussian function, we indicate each bin by means of three data points, at the start, center, and the end of it, see Fig.~\ref{fig:Gaussian fit}. Each data point indicates the value and the associated uncertainty of the corresponding bin. Fitting a Gaussian model on the aforementioned distribution yields the $\sigma$ and its uncertainty, which we adopt as the size of the cluster and its uncertainty.} Hence, the cluster has $10^{+3}_{-2}$ members within the core region \footnote{Eight sources located in the area with radius of $18\,{\rm mpc}$ : N6, 12, 15, 19, 23, 24, 25, 26. \\ Ten sources located in the area with the radius of $22\,{\rm mpc}$ : N5, 6, 7, 12, 15, 19, 23, 24, 25, 26. \\ 13 sources located in the area with the radius of $25\,{\rm mpc}$ : N5, 6, 7, 11, 12, 15, 18, 19, 22, 23, 24, 25, 26.}.
We determine the required binding mass by applying the virial theorem, $M_{\rm IMBH}\sim \sigma_{\star}^2 R_{\rm c}/G$, where $R_{\rm c}=23 \pm 4\,{\rm mpc}$, $G$ is the gravitational constant, and $\sigma_{\star}$ is the stellar velocity dispersion. Hence, we assume that the NW-source cluster is relaxed, which may not be the case in general. For the calculation of the virial binding mass, we consider just sources whose total on-sky proper motion is below the escape velocity at a given projected distance from Sgr~A* (at $\sim 6.245''\sim 0.25\,{\rm pc}$, the escape velocity is $\sim 370\,{\rm km\,s^{-1}}$). This way we remove eight sources (N6, N10, N18, N24, N26, N39, N40, N42) that exceed this value, which is caused by confusion due to their faintness.

For the remaining 20 NW sources, we consider two perpendicular directions for the estimate of $\sigma_{\star}$ -- along the right ascension and the declination, each corrected for the corresponding mean velocity error. For $\sigma_{\rm RA} = 58.9^{+1.5}_{-0.7}\,{\rm km\,s^{-1}}$, we obtain $M_{\rm IMBH} =18.6^{+0.3}_{-0.3} \times 10^3 M_{\odot}$ and for  $\sigma_{\rm Dec} = 64.8^{+2.6}_{-1.6}\,{\rm km\,s^{-1}}$, $M_{\rm IMBH}= 22.5^{+0.4}_{-0.4}\times 10^3 M_\odot$\footnote{These velocity dispersion values are changed by a few ${\rm km\,s^{-1}}$ ($\sigma_{\rm RA}=62.3^{+1.6}_{-0.8}\,{\rm km\,s^{-1}}$, $\sigma_{\rm Dec}=63.9^{+1.9}_{-0.9}\,{\rm km\,s^{-1}}$) by removing two potential late-type stars N9 and N30, which does not impact the IMBH virial mass estimate significantly.}. Thus, using the virial theorem, we obtain the first estimate of the required IMBH mass of the order of $\sim 10^4\,M_{\odot}$ to bind the NW-ward flying sources gravitationally. The second estimate comes from the tidal stability criterion of the putative NW-ward cluster. This can be formulated using the condition that the effective cluster radius is smaller or comparable to its tidal (Hill) radius as it orbits Sgr~A* SMBH, hence
\begin{equation}
    R_{\rm c}\lesssim d_{\rm NW}\left(\frac{m_{\rm NW}}{3M_{\bullet}} \right)^{1/3}\,,
    \label{eq_tidal_radius}
\end{equation}
where $d_{\rm NW}\gtrsim 6.05''\sim 0.242\,{\rm pc}$ is the distance of the NW-ward flying association from Sgr~A*, where the lower limit is given by the projected distance. For the SMBH mass, we take $M_{\bullet}\sim 4\times 10^6\,M_{\odot}$ and the cluster radius is estimated from the dispersion of the Gaussian fit to the number density distribution, $R_{\rm c}=23 \pm 4\,{\rm mpc}$ as before. From the tidal stability condition given by Eq.~\eqref{eq_tidal_radius}, the total mass of the NW-ward cluster then is $m_{\rm NW}\gtrsim 3M_{\bullet}(R_{\rm c}/d_{\rm NW})^3$, which gives $m_{\rm NW}\gtrsim (10.3 \pm 5.4)\times 10^3\,M_{\odot}$. Considering the total radius of the N-source region, $R_{\rm N}\sim 1.35''\sim 54\,{\rm mpc}$, we obtain $m_{\rm NW}\gtrsim 1.33\times 10^5\,M_{\odot}$. Since the stars contribute $\lesssim 10^3\,M_{\odot}$ to the cluster mass, which follows from the total number of detected stars of the order of 10, the required mass for the tidal stability of the cluster is significantly larger and could be complemented by an IMBH of the mass $m_{\rm IMBH}\sim m_{\rm NW}\sim 10^4-10^5\,M_{\odot}$, considering the total range of the N-source region size. The mass range $\sim 10^4-10^5\,M_{\odot}$ of the hypothetical IMBH associated with the NW-ward flying association, which was inferred using both the virial theorem and the tidal stability criterion, is consistent with the expected IMBH mass of $\sim 10^2-10^5\,M_{\odot}$ in various stellar environments \citep{2020ARA&A..58..257G}.  



\begin{figure}[tbh!]
\begin{center}
 \includegraphics[width=0.5\textwidth]{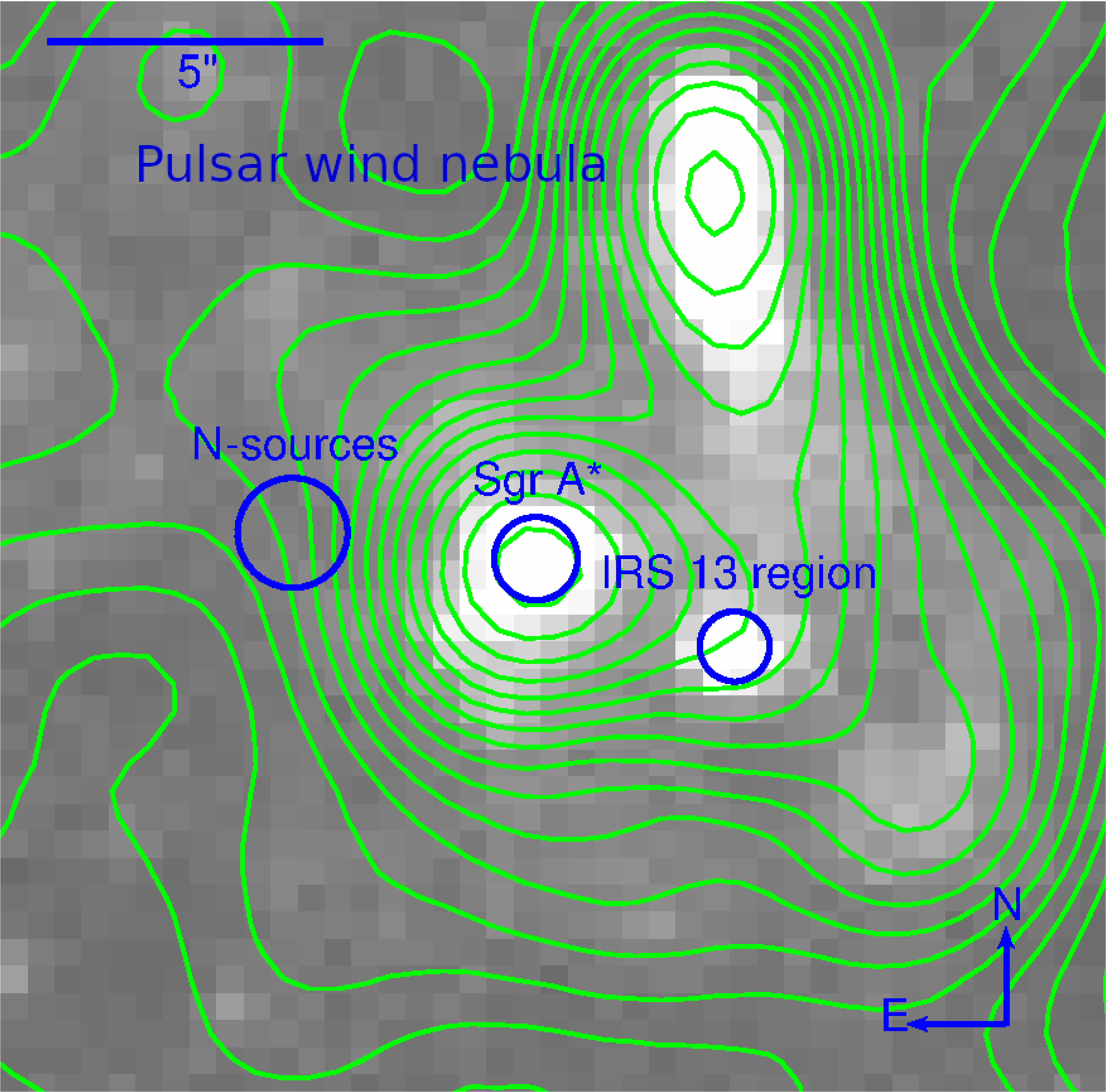}
   \caption{The \textit{Chandra} X-ray image from 0.5 to 8 keV. The north is up and the east to the left. Distinct X-ray bright sources are labelled, namely Sgr~A*, IRS 13, and the pulsar wind nebula. There is no significant X-ray source at the position of the N-source association. The angular scale is indicated in the upper left corner.}
\label{Fig:x-ray 0.5-8}
\end{center}
\end{figure}

Using the 1999–2012 X-ray \textit{Chandra} data available from the \textit{Chandra} Search and Retrieval interface, we limit our study to the observations
when Sgr~A* was observed with an off-axis angle lower than 2 arcmin.
This results in 84 observations with the ACIS-I or ACIS-S/HETG cameras
\citep{2003SPIE.4851...28G} and a total exposure time of 4.6 Ms (see details in \cite{Mossoux2017}). The pixel scale of the \textit{Chandra} X-ray data is 492 ${\rm mas}/$pixel. In Fig.~\ref{Fig:x-ray 0.5-8} and Fig.~\ref{Fig:x-ray} (Appendix), we plot contours at 100, 80, 60, 40 \% of the peak counts detected at the position of Sgr~A*. In the central few arcseconds, \textit{Chandra} observations have revealed
both the extended X-ray emission as well as the emission from compact sources
that can be associated with stars (e.g. X-ray binaries).
These data confirm a thermal
X-ray spectrum of IRS13E. Fitting the spectrum with an
optically thin plasma model, one finds a temperature of $2.0$ keV
and an unabsorbed $2$–$10$ keV luminosity
of about $\sim 2.0 \times 10^{33} {\rm erg s^{-1}}$ for the IRS13 region \citep{Zhu2020, Wang2006}.

No X-ray source could be identified at the position of N-sources.
We find that in a $1.5''$ diameter aperture the X-ray luminosity at
the position of N-sources is about three times fainter than that of IRS 13
in the $0.5$ to $8~keV$ image and about 10 times fainter in the 4 to $8~keV$ image.
Hence, the X-ray luminosity that can be associated with the N-source region is below
$L_{\rm X}\lesssim 0.2 \times 10^{33} {\rm erg s^{-1}}$.
In comparison to IRS 13E, this may indicate either a lack of
hot optically-thin plasma or an increased extinction from the foreground material,
possibly associated with the circumnuclear disk
(CND, see \citeauthor{Mossoux2017}, \citeyear{Mossoux2017}). In fact, \citet{2017A&A...603A..68M} report high-velocity SiO gas south-west of IRS 1W, see their Fig.~11. 

 However, the near-infrared extinction maps \cite[][]{2010A&A...511A..18S,2016ApJ...821...44F,2016A&A...594A.113C}
do not support an excessively higher extinction towards that region
on the arcsecond scale.
On the other hand, it is unclear how the N-sources in the field are
located with respect to the northern arm in comparison to what
Fig.\ 8 of \citet{2020ApJ...896...68N} indicates. Therefore, a dearth of hot,
optically-thin plasma is also not a clear cut explanation. Hence,
to explain the lack of X-ray emission towards the N-sources
we are left with either a combination of the above two effects and/or
an intrinsically weak X-ray emission from that region as such.

The lack of X-ray emission does not support the
presence of a hypothetical IMBH accreting in a radiatively efficient way,
see e.g. \cite{Schoedel2005}. In other words, the Eddington ratio\footnote{The Eddington luminosity is estimated using the standard upper limit for the steady spherical accretion, $L_{\rm Edd}=4\pi G m_{\rm IMBH} m_{\rm p} c/\sigma_{\rm T}$.} of the system $\lambda_{\rm Edd}=\kappa_{\rm bol} L_{\rm X}/L_{\rm Edd}\lesssim 3.4 \times 10^{-6} \left(\frac{L_X}{2\times 10^{32}\,{\rm erg\,s^{-1}}} \right)\left(\frac{m_{\rm IMBH}}{10^4\,M_{\odot}} \right)^{-1}$, where we used the bolometric correction $\kappa_{\rm bol}=13\left(\frac{L_{\rm X}}{10^{41}\,{\rm erg\,s^{-1}}} \right)^{-0.37}$ according to \citet{2014MNRAS.438.2804N}, would imply a radiatively inefficient mode of accretion in the form of e.g. Advection Dominated Accretion Flow (ADAF), which is expected for IMBHs across various galactic environments \citep[see e.g.][]{2022MNRAS.515.2110S,2022arXiv221100019P}. Under the approximation of the spherical, Bondi-like accretion of the surrounding hot plasma onto the IMBH, we can estimate the Bondi rate as 
\begin{equation}
    \dot{M}_{\rm B}=\frac{4\pi G^2 m_{\rm IMBH}^2\mu n_{\rm a}m_{\rm H}}{c_{\rm s}^3}\,,
    \label{eq_Bondi_accretion}
\end{equation}
where $n_{\rm a}$ is the ambient medium number density, $c_{\rm s}$ is the sound speed in the ambient plasma, $m_{\rm H}$ is the hydrogen atom mass, and $\mu$ is the mean molecular weight ($\mu\sim 0.5$ for fully ionized plasma). As an approximation, we first consider the ambient plasma properties similar to those inferred for the Bondi radius, i.e. $n_{\rm a}\sim 26\,{\rm cm^{-3}}$ and $kT_{\rm a}\sim 1.3\,{\rm keV}$ \citep{2003ApJ...591..891B}. For $m_{\rm IMBH}=10^4\,M_{\odot}$, we obtain $ \dot{M}_{\rm B}\sim 6.2\times 10^{-11}\,M_{\odot}{\rm yr^{-1}}$ according to Eq.~\eqref{eq_Bondi_accretion}. Using the relation for the bolometric luminosity, $L_{\rm bol}= \kappa_{\rm bol}L_{\rm X}\sim 4.3 \times 10^{36} (L_X/2\times 10^{32}\,{\rm erg\,s^{-1}})^{0.63}$ and assuming the radiative efficiency of $\eta_{\rm R}\sim 0.1$, the actual X-ray luminosity is at most $L_{X}\lesssim 3.8 \times 10^{30}\,{\rm erg\,s^{-1}}$ since the constraints on the density and the radiative efficiency are the upper limits. Hence, Bondi-like hot flow feeding the putative IMBH can explain its low X-ray emission and thus the difficulty to detect it directly within the nuclear star cluster, not only in the IRS 1W NW-ward flying region, but also in IRS 13E.

In the more precise accretion model, we attempt to construct the spectral energy distribution (SED) of the hot ADAF surrounding the putative IMBH of $m_{\rm IMBH}\sim 10^4\,M_{\odot}$. The basic input parameters of the ADAF SED are the IMBH mass $m_{\rm IMBH}$ and its relative accretion rate $\dot{m}\equiv \dot{M}_{\rm acc}/\dot{M}_{\rm Edd}$ \citep{2021ApJ...923..260P}, where $\dot{M}_{\rm acc}$ is the actual accretion rate of the IMBH at the inner radius, which is a fraction of the Bondi accretion rate $\dot{M}_{\rm B}$, and $\dot{M}_{\rm Edd}=L_{\rm Edd}/(\eta_{\rm R}c^2)$ is Eddington accretion rate. Since for the adopted $m_{\rm IMBH}=10^4\,M_{\odot}$ we can estimate $\dot{M}_{\rm Edd}$ in a straightforward way, we need to better estimate $\dot{M}_{\rm acc}$ from $\dot{M}_{\rm B}$, see Eq.~\eqref{eq_Bondi_accretion}. The sound speed of the flow can be estimated from the virial theorem, assuming that the gas and the stellar motion are virialized. Then we can associate the mean quadratic velocity of the gas $<v^2>$ with the stellar velocity dispersion, $<v^2>=\sigma_{\star}^2=3\sigma_1^2$, where $\sigma_1$ is a one-dimensional dispersion along any direction. The average one-dimensional dispersion from RA and Dec. dispersions is $\sigma_i\sim 62\,{\rm km/s}$. Then we can estimate the sound speed in the gas as $c_{\rm s}=<v^2>^{1/2}/\sqrt{3}\sim \sigma_i\sim 62\,{\rm km/s}$, which corresponds to the temperature of $T_{\rm a}\sim 2.3\times 10^5\,{\rm K}$ or $\sim 19.9\,{\rm eV}$. The inferred sound speed of the gas being captured by the IMBH then yields the estimate of the capture or the Bondi-radius,
\begin{align}
    R_{\rm B}&=\frac{2Gm_{\rm IMBH}}{c_{\rm s}^2}\,\notag\\
    &=22.5\left(\frac{m_{\rm IMBH}}{10^4\,M_{\odot}} \right)\left(\frac{T_{\rm a}}{2.3\times 10^5\,{\rm K}} \right)^{-1}{\rm mpc}\sim 0.56''\,,
    \label{eq_bondi_radius}
\end{align}
which is comparable to the cluster core radius $R_{\rm c}$.

\begin{figure}
    \centering
    \includegraphics[width=\columnwidth]{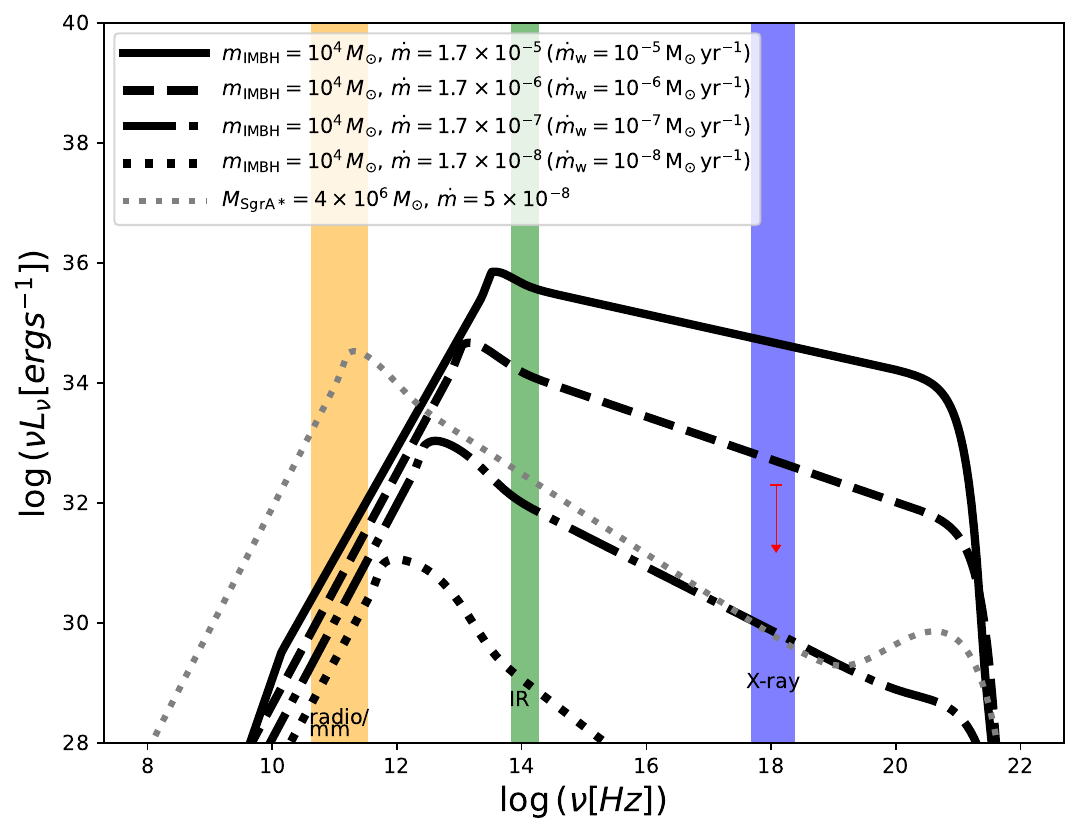}
    \caption{SEDs of advection-dominated accretion flows associated with the putative IMBH at the geometrical center of the NW-ward flying sources. Different black lines stand for the accretion rates according to the legend. For comparison, we also show the SED of Sgr~A* (gray dotted line). The red point depicts the X-ray emission upper limit.}
    \label{fig_IMBH_SED}
\end{figure}

For the estimate of the actual accretion rate $\dot{M}_{\rm acc}$, we use the power-law relation between the accretion rate at the Bondi radius $R_{\rm B}$ and the inner radius, which we set to $R_{\rm in}=100\,Gm_{\rm IMBH}/c^2$ following \citet{2022MNRAS.515.2110S} and references therein,
\begin{equation}
    \dot{M}_{\rm acc}=\dot{M}_{\rm B}\left(\frac{R_{\rm in}}{R_{\rm B}} \right)^p
    \label{eq_accretion_rate}
\end{equation}
where the power-law index is set to $p=0.5$. To estimate the mean density of the ambient flow $n_{\rm a}$, we assume a set-up where $N_{\star}$ stars are uniformly distributed in the sphere of radius $R_{\rm c}$, which represents the cluster core radius. Then the mean number density can be inferred from the stationary wind outflow at the mean distance between the star and the IMBH or
\begin{align}
    n_{\rm a} &= \frac{N_{\star}\dot{m}_{\rm w}}{4 \pi \overline{r}^2 \mu m_{\rm H}v_{\rm w}}\,\notag\\
    &=\left(\frac{3N_{\star}}{4\pi} \right)^{5/3} \frac{\dot{m}_{\rm w}}{v_{\rm w}} (3 \mu m_{\rm H}R_{\rm c}^2)^{-1}\,,
    \label{eq_number density}
\end{align}
where $\dot{m}_{\rm w}$ is the mean mass-loss rate per star and $v_{\rm w}$ is the terminal stellar-wind velocity. Combining Eqs.~\eqref{eq_Bondi_accretion}, \eqref{eq_accretion_rate}, and \eqref{eq_number density}, along with the inferred sound speed, we obtain the Eddington ratio of the IMBH of $\dot{m}=\dot{M}_{\rm acc}/\dot{M}_{\rm Edd}\sim 1.7 \times 10^{-5}$ for the upper limit of the stellar mass-loss rate ($\dot{m}_{\rm w}=10^{-5}\,{\rm M_{\odot}\,yr^{-1}}$, $v_{\rm w}=10^3\,{\rm km/s}$) and $\dot{m}=1.7 \times 10^{-8}$\, for a lower stellar mass-loss rate of $\dot{m}_{\rm w}=10^{-8}\,{\rm M_{\odot}\,yr^{-1}}$ ($v_{\rm w}=10^3\,{\rm km/s}$). This range of relative accretion rates of the putative IMBH at the core of the NW-cluster can be used to calculate the SED of the corresponding ADAF \citep{2021ApJ...923..260P} and hence the observability of the candidate IMBH across different wavelengths. In Fig.~\ref{fig_IMBH_SED}, we show ADAF SEDs that were calculated using the code \texttt{LLAGNSED} \citep{2021ApJ...923..260P} for different accretion rates (see the legend). For decreasing accretion rates, SED peaks shift towards longer wavelengths from the infrared towards the radio/mm domain. For comparison, we also depict the SED of Sgr~A*, which peaks in the mm domain. Using the X-ray upper limit of $\leq 0.2 \times 10^{33}\,{\rm erg\,s^{-1}}$, we can constrain the Eddington ratio to $\dot{m} < 1.7\times 10^{-6}$, for which the feeding by $N_{\star}=10$ stars with $\dot{m}_{\rm w}\lesssim 10^{-6}\,{\rm M_{\odot}\,yr^{-1}}$ within the cluster core is sufficient.

In case there is no IMBH at the center of the NW-ward flying association, it will be tidally disrupted on the orbital timescale,
\begin{align}
    P_{\rm orb}&\gtrsim 2\pi \frac{d_{\rm NW}^{3/2}}{\sqrt{GM_{\bullet}}}\,\notag\\
    &\sim 5577\,\left(\frac{d_{\rm NW}}{0.242\,{\rm pc}} \right)^{3/2}\left(\frac{M_{\bullet}}{4\times 10^6\,M_{\odot}} \right)^{-1/2}{\rm yr}\,,
\end{align}
which is such a short timescale in comparison with the stellar lifetime that it is unlikely that we observe the NW-ward group just at the beginning of tidal dissociation. Therefore, it is more likely that either the group is bound by a quiescent IMBH or the overdensity is due to the projection effect, in particular the inclined disk-like distribution of stars that we discuss in the following subsection.  

Eventually, the lifetime of the IMBH-bound stellar association similar to the NW-ward flying group is given by the dynamical friction timescale, during which the IMBH inspirals towards Sgr~A*,
\begin{align}
    \tau_{\rm df}&=\frac{3}{8}\sqrt{\frac{2}{\pi}}\frac{\sigma_{\star}^3}{G^2\rho_{\star}m_{\rm IMBH}\rm{ln}\Lambda}\,\notag\\
    &\sim 3\times 10^5 \left( \frac{\sigma_{\star}}{88\,{\rm km\,s^{-1}}}\right)^3\left(\frac{\rho_{\star}}{2.4\times 10^5\,M_{\odot}\,{\rm pc^{-3}}} \right)^{-1} \notag\\
    &\times \left( \frac{m_{\rm IMBH}}{10^4\,M_{\odot}}\right)^{-1} \left(\frac{{\rm ln}\Lambda}{15} \right)^{-1} \,{\rm yr}\,,
    \label{eq_df_timescale}
\end{align}
where $\sigma_{\star}$ is the stellar velocity dispersion that we scale to $\sigma_{\star}\sim 88\,{\rm km\,s^{-1}}$ according to $M_{\bullet}$-$\sigma_{\star}$ relation \citep{2009ApJ...698..198G}; the stellar mass density $\rho_{\star}$ is estimated using the condition that there is the stellar mass of $M_{\star}(r<r_{\rm h})\sim 2\,M_{\bullet}$ inside the Sgr~A* sphere of influence, which is $r_{\rm h}\sim 2\,{\rm pc}$ \citep{2013degn.book.....M}; the IMBH mass is scaled to $m_{\rm IMBH}\sim 10^4\,M_{\odot}$ following the previous estimates, and the Coulomb logarithm is estimated as $\rm{ln}\Lambda\sim \rm{ln}(M_{\bullet}/M_{\odot})\sim 15$. Hence, the stellar association would be disrupted on the timescale of the order of $\tau_{\rm df}$ as the IMBH inspirals via dynamical scattering through the nuclear star cluster. The tidal stripping of the stellar cluster during the IMBH inspiral towards the SMBH may be one of the mechanisms how early-type stars of spectral type O and B are deposited close to the SMBH on the scale of $\sim 0.01$ pc.


\subsection{Disk-like distribution projection}

\label{section:Disk projection}

\begin{figure}
\begin{center}
 \includegraphics[width=\linewidth]{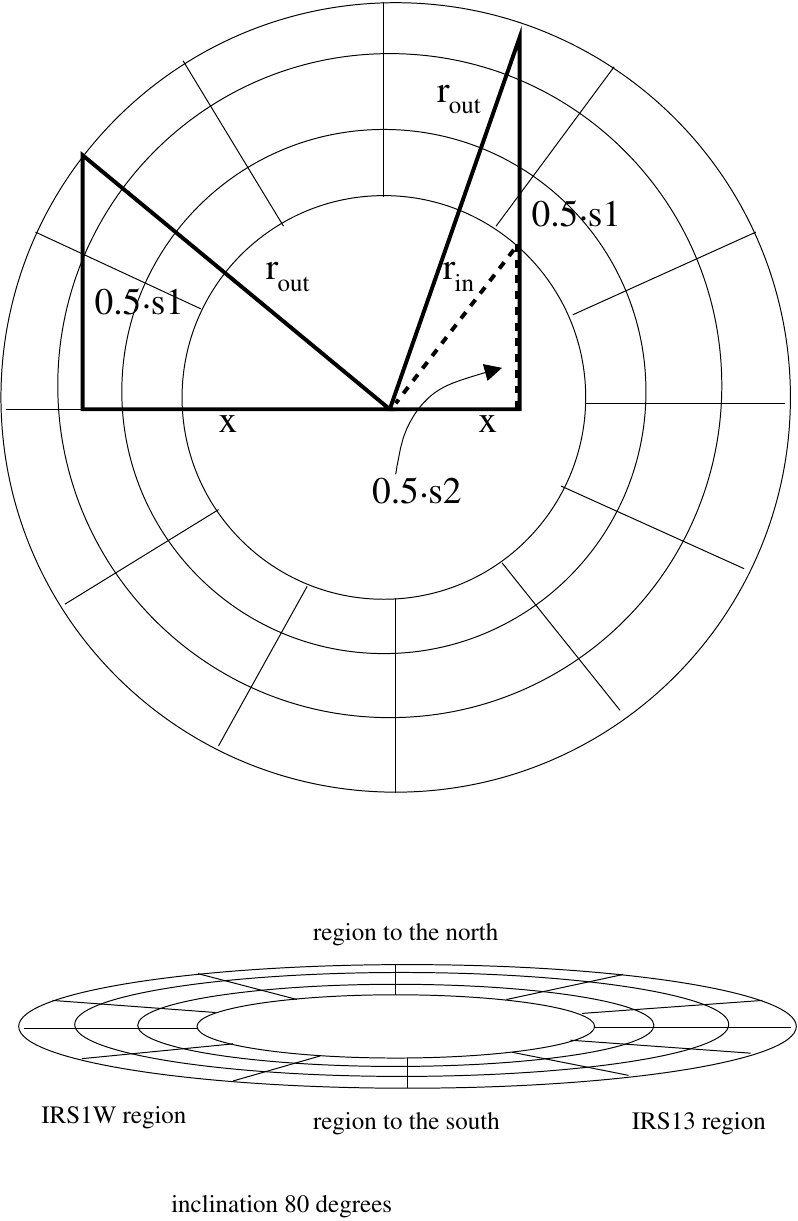}
   \caption{Top: Face on view of the East West (EW) stellar disk system with outer and inner radii given in the text. For the case of $s1$ we indicated the corresponding quantities. Bottom: The disk system is inclined by 80 degrees. The clustering at the eastern and western tips of the disk become evident. }
\label{fig:projection}
\end{center}
\end{figure}

\begin{figure}
\begin{center}
\includegraphics[width=\columnwidth]{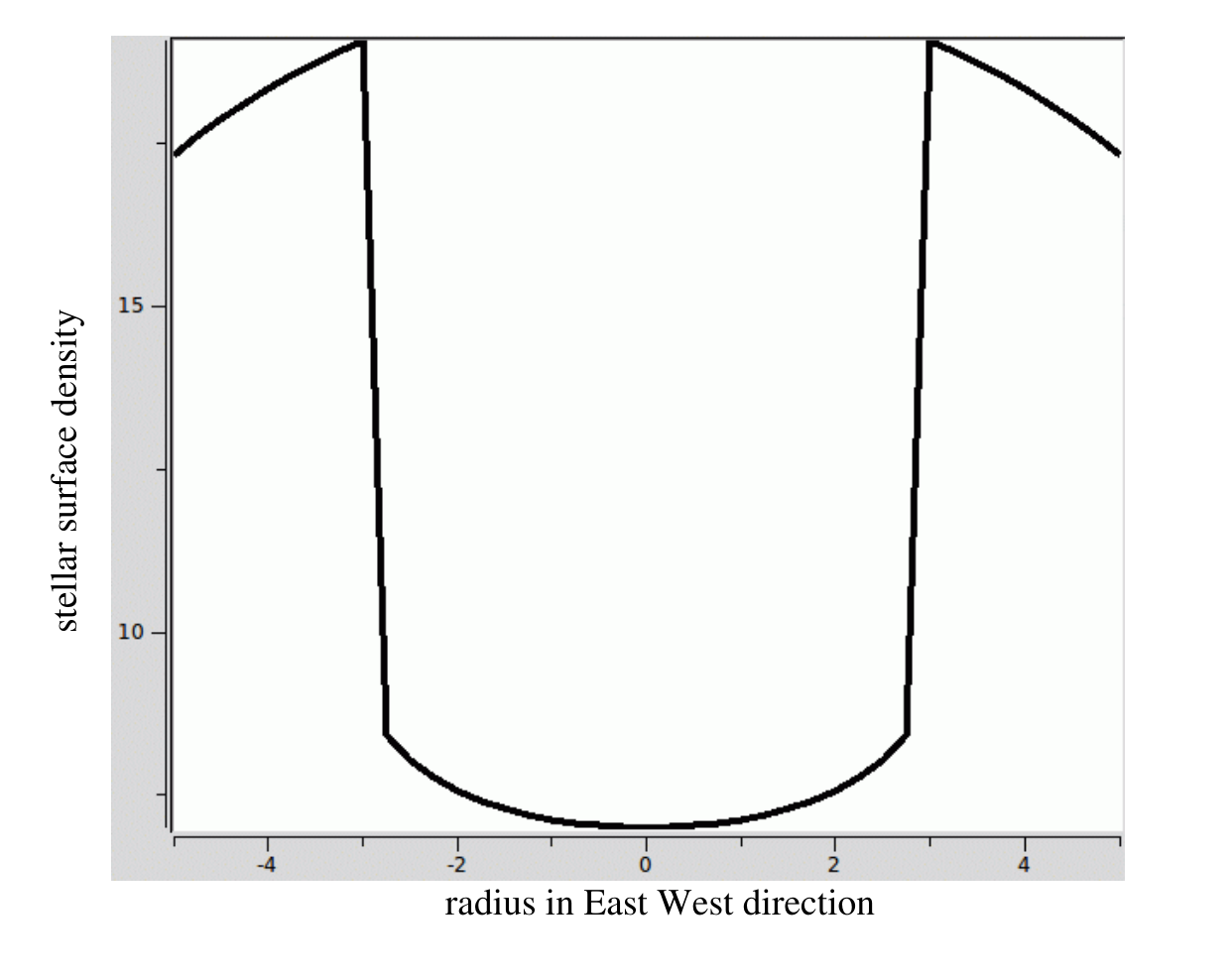}
\caption{Stellar sky surface density in arbitrary units as a 
function of radius in the east west direction in arcseconds. 
Here also, the clustering at the eastern and western
tips of the disk become evident. The source density is almost a factor of 2 higher than in the northern and southern part of the disk.}
\label{fig:density}
\end{center}
\end{figure}

\citet{Paumard2006} and \citet{Lu2009} categorized IRS 1W as a member of the clockwise disk and IRS13E members were introduced as members of the counter-clockwise disk. However, later \cite{Yelda2014} did not find evidence for the existence of the counter-clockwise disk and \citet {Sanchez-Bermudez2014} also did not categorize IRS 1W as a member of the clockwise disk based on the bow-shock orientation and the proper motion. The fact that the N-sources and the IRS13 region are located on opposite sides to each other at similar distances from Sgr A* ($6''$ and $4''$, respectively), suggests that their appearance may in fact be the result of the projection of a disk-like distribution, such as the clockwise disk of young He-stars and/or dust-enshrouded stars.

Looking at the proper motion velocities of individual sources
under the assumption of the disk scenario, one has to bear in mind that
this is not a very stringent picture. We are not dealing with a
thin disk with orbiting stars fixed to it. Rather the disk has a
non-vanishing thickness, i.e., there is a certain velocity dispersion in the direction
 perpendicular to the disk plane. In addition, the disk stars are on individual
orbits with varying orbital elements (i.e.\ the orbit orientation and
eccentricity). Hence, the current data does not allow us to make
firm statements on their disk deprojected motion.

We show that the corresponding projected stellar surface density on the sky may be about a factor of two higher than the density of the northern and southern projected disk sections.
We can estimate this effect in the following way.
We assume that the disk-forming fraction of stars is arranged in
a disk with an inclination of $i = 10$ degrees \citep{Levin2003, Lu2008}, with an outer radius of $r_{\rm out}=5''$, and an inner radius of $r_{\rm in}=1.5''$ (shown in Fig. \ref{fig:projection}). We can then calculate cross-sections through that disk system as:
\begin{equation}
\mathit{s_1}=2\sqrt{\lvert r_{\rm out}^2 - x^2\lvert}
\end{equation}

        (see the left side of the upper panel in Fig. \ref{fig:projection}.)

\begin{equation}
s_2=s_1-2\sqrt{\lvert r_{\rm in}^2 - x^2\lvert}
\end{equation}
       \par (see the right side of the upper panel in Fig. \ref{fig:projection}.)\\
       
The projected surface density $\rho$ is proportional to $s_1/\cos(i)$ for 
the regions beyond the inner radius $r_{\rm in}$
and is proportional to $\frac{1}{2}s_2/\cos(i)$ for 
the regions within the inner radius $r_{\rm in}$ to the north and the south of Sgr~A*.
The fact that both N-sources and IRS13 lie on the opposite sides of Sgr~A* at
a rather similar distance may support this model.
Hence, the overdensity of these compact stellar systems could be explained by the projection effect in which the excess in velocity dispersion is given to a large part by the disk dynamics, see Fig.\ref{fig:density} .

\section{Conclusions}
\label{section:Conclusion}

For the first time, we show that N-sources are north-ward moving sources located in the Galactic Center in the vicinity of the bow-shock source IRS 1W. We demonstrate that these sources possess a Gaussian-like number density distribution with respect to the geometrical center. The overdensity is apparent in comparison with the number of sources in random test regions at a comparable distance from Sgr~A*. These circular regions have the same length-scale as the N-source region. The mean number of the detected sources in our eleven test regions is $14.6 \pm 1.5$, while the number of N-sources is 42, which indicates that the N-source group is potentially a stellar association in the vicinity of Sgr~A*. 

Further investigation of the proper motions of N-sources reveals that N-sources can be divided into two categories. One category encompasses the NW-ward flying sources which show a spatial Gaussian number density distribution, while the NE-ward flying sources have a rather spatially flat distribution. Therefore, the NW-ward flying sources could potentially be bound or have a common origin, while NE-ward flying sources most likely consist of local background/foreground stars that are part of the nuclear star cluster.

The high concentration of the N-sources, mainly the NW-ward flying group, can be interpreted in terms of the stellar cluster that is kept stable by the central IMBH or it could be the result of a disk-like stellar distribution that is projected at a high angle. 

The first scenario is supported by a likely occurrence of IMBHs in the Galactic center. One possible formation channel is a series of collisions of stellar-mass black holes with main-sequence stars, which was analyzed by \citet{Rose2022}.  This mechanism appears quite efficient and it can produce IMBHs of mass $\lesssim 10^4\,M_{\odot}$, which is within the current uncertainties of the IMBH mass constraints for IRS 1W association, specifically using either the velocity dispersion of NW-ward flying sources or the tidal stability criterion. 

Another potential way is the cluster infall scenario from larger scales \citep[see e.g.][and references therein]{2022arXiv220205618F}, which assumes that massive clusters, such as young star-forming clusters or globular cluster host IMBHs with a certain occupation fraction. In this regard, the NW-ward flying group could be a remnant, tidally stable core of such an infalling massive cluster. Using the \textit{Chandra} X-ray telescope data, we show that such an IMBH would have a low Eddington ratio of $\dot{m}< 1.7 \times 10^{-6}$, hence would be surrounded by an X-ray faint advection-dominated flow. Future spectroscopic data collected by the James Webb Space Telescope and 30/40-meter class telescopes will be crucial for a better understanding of the stellar composition of the N-sources, and this may also help to clarify their potential origin in a common cluster whose core can only be kept stable by an IMBH at a given distance from Sgr~A*. On the other hand, the short dynamical friction timescale of an IMBH within the NSC of the order of $10^5$ years, 
 see Eq.~\eqref{eq_df_timescale}, implies that the IMBH, including its stellar system, is a dynamically transient phenomenon in the dense environment of the NSC, which also applies to the IRS 13 association.

The second scenario is supported by an exceptional distribution of cluster-like stellar associations in the vicinity of Sgr~A*. Specifically, the IRS13 and the N-source areas are positioned at $\sim 4''$ south-west and $\sim 6''$ north-east of Sgr~A*, respectively, nearly along the same line, which raises the possibility that the observed overdensities may actually be formed due to the projection when the two regions are situated on the same stellar disk with the characteristics similar to that of the clockwise stellar disk. 

In the current investigation, we do not find a strong preference for any of the two models. It is therefore necessary to conduct detailed spectroscopic measurements to rule out one of the models. More information about the orbital properties of these sources, in particular spectroscopically determined line-of-sight velocities, will be crucial for the further study of this and similar regions.

\section*{Acknowledgments}
\begin{acknowledgements}
  We thank an anonymous referee for constructive comments that helped to improve the manuscript.
  This work was supported in part by SFB 956—Conditions and Impact of Star Formation. S. Elaheh Hosseini is a member of the International Max Planck Research School for Astronomy and Astrophysics at the Universities of Bonn and Cologne. We thank the Collaborative Research Centre 956, sub-project A02, funded by the Deutsche Forschungsgemeinschaft (DFG) – project ID 184018867. M.Z. acknowledges the financial support by the GA\v{C}R Junior Star No. GM24-10599M ``Stars in galactic nuclei: interrelation with massive black holes". V.K. acknowledges Czech Science Foundation grant No.\ 21-06825X and the Czech Ministry of Education, Youth and Sports Research Infrastructure No.\ LM2023047.
\end{acknowledgements}

\appendix

\begin{figure}[htbp!]  
\begin{center}
\includegraphics[width=0.5\linewidth]{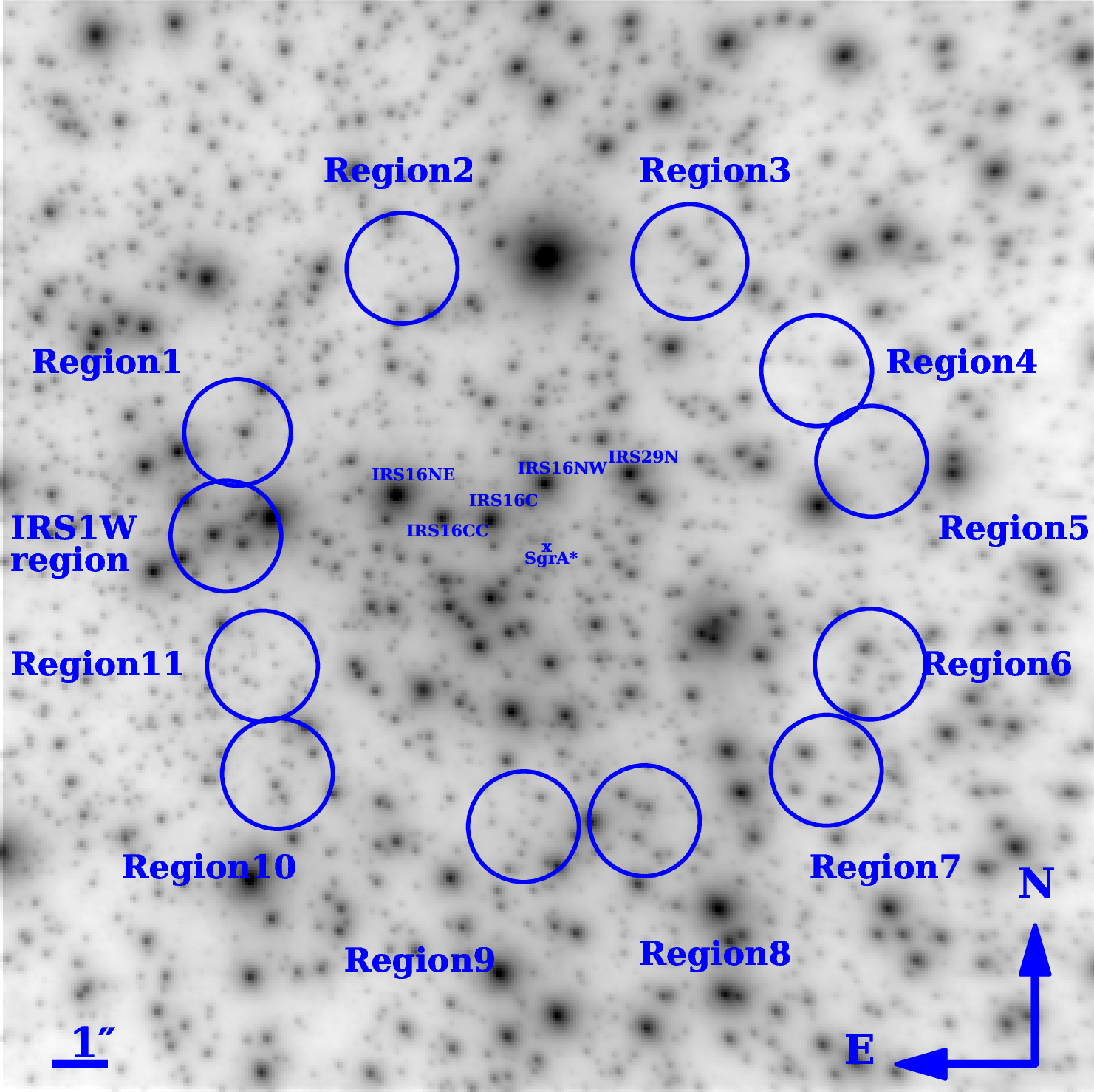}
\caption{The $K_s$-band from 2005.366 observation. The $K_s$-band image denotes eleven regions in addition to the assumed region of IRS 1W region. These eleven regions are chosen randomly in order to identify the number of sources in each one. The only criterion is the distance from Sgr~A*; all regions are located in the distance of almost 6\,$''$ from Sgr~A*. The radii of all circular regions are about 1.3\,$''$.}
\label{fig:11 regions}
\end{center}
\end{figure}

\begin{figure}
     \centering
     \subfloat{\includegraphics[width=0.48\textwidth]{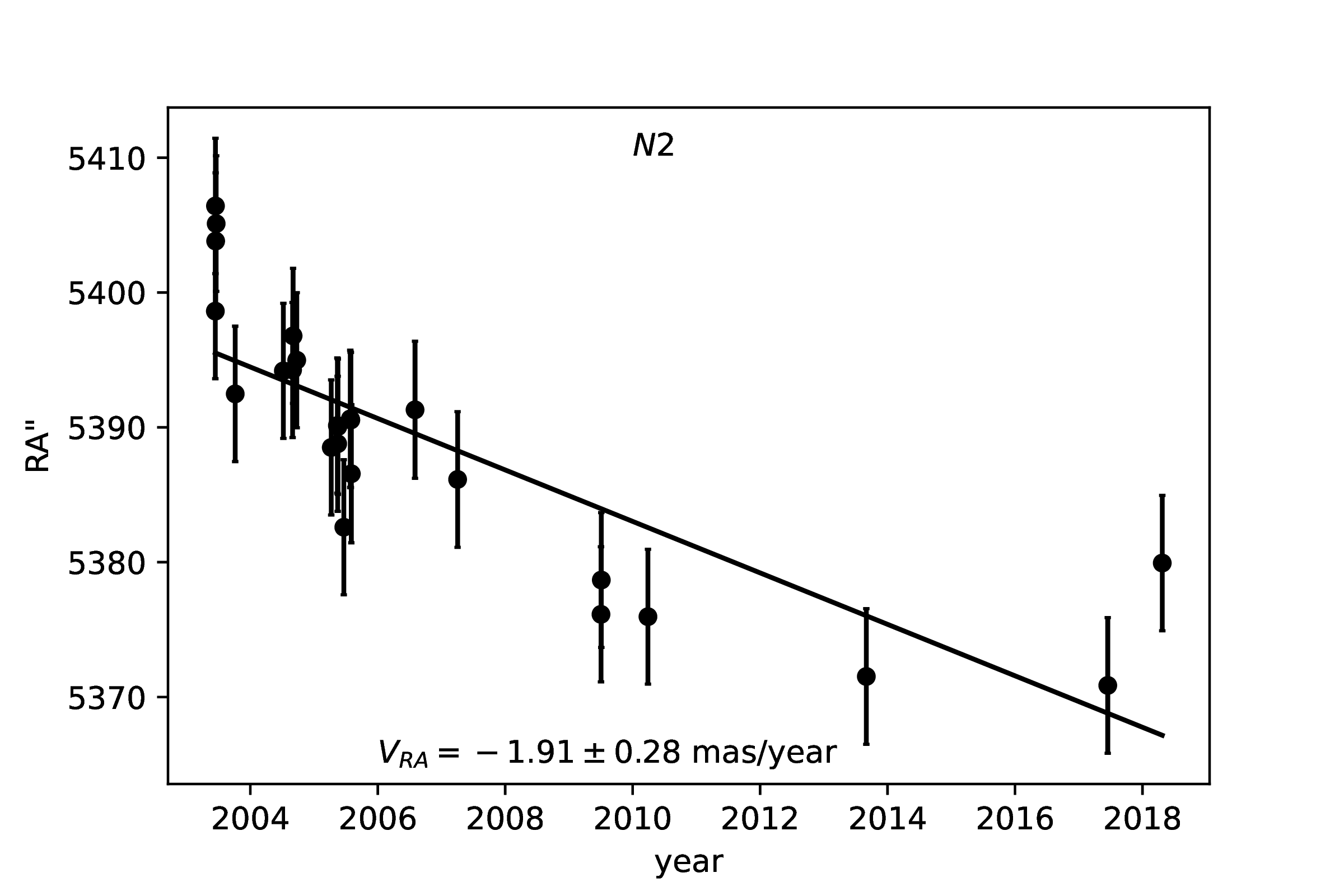}}
     \subfloat{\includegraphics[width=0.48\textwidth]{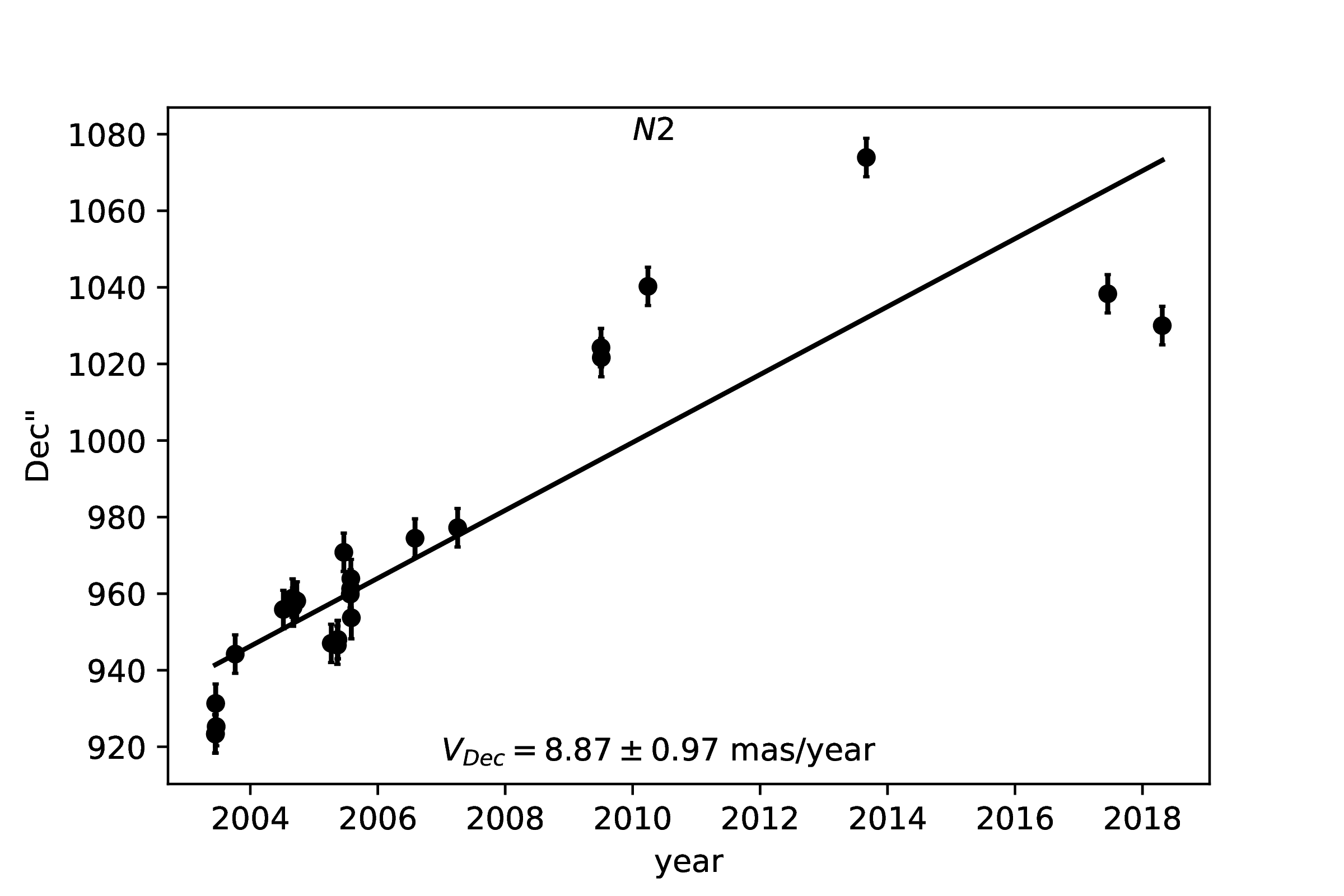}}

     \caption{The plots show the derived positions of N2 star as a function of time, along with the best-fitting proper motions. The left panel is along RA and the right panel is along Dec. The slope of each best-fitting line is given in Table \ref{table:proper motion wsources km/s} (in {\it km/s}).}
     \label{plot:proper motion of w2}
 \end{figure}

\begin{figure}[tbh!]
\begin{center}
 \includegraphics[width=0.5\textwidth]{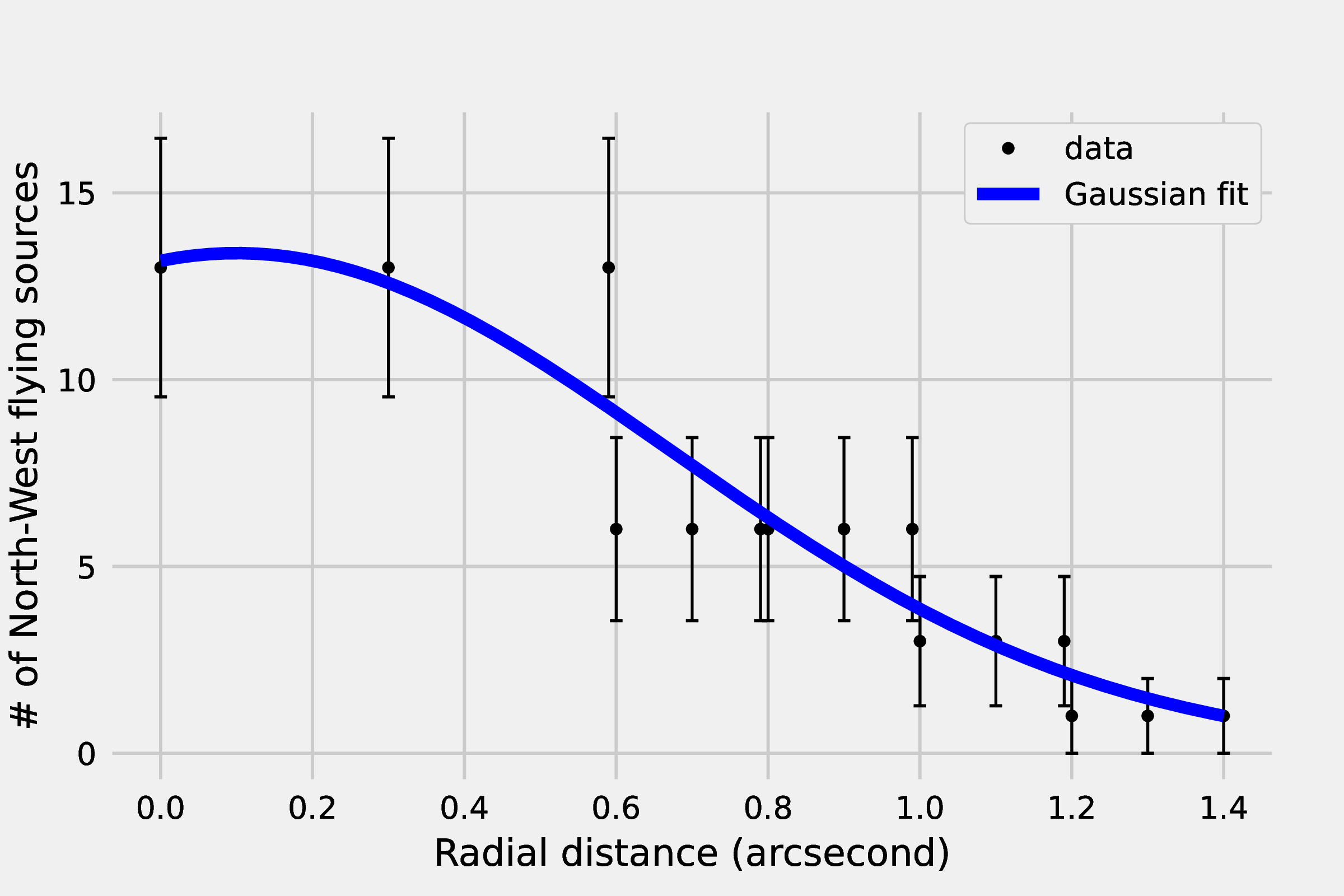}
   \caption{Gaussian fit of the spatial distribution of NW-flying sources.}
\label{fig:Gaussian fit}
\end{center}
\end{figure}

\begin{figure}
\begin{center}
 \includegraphics[width=0.5\textwidth]{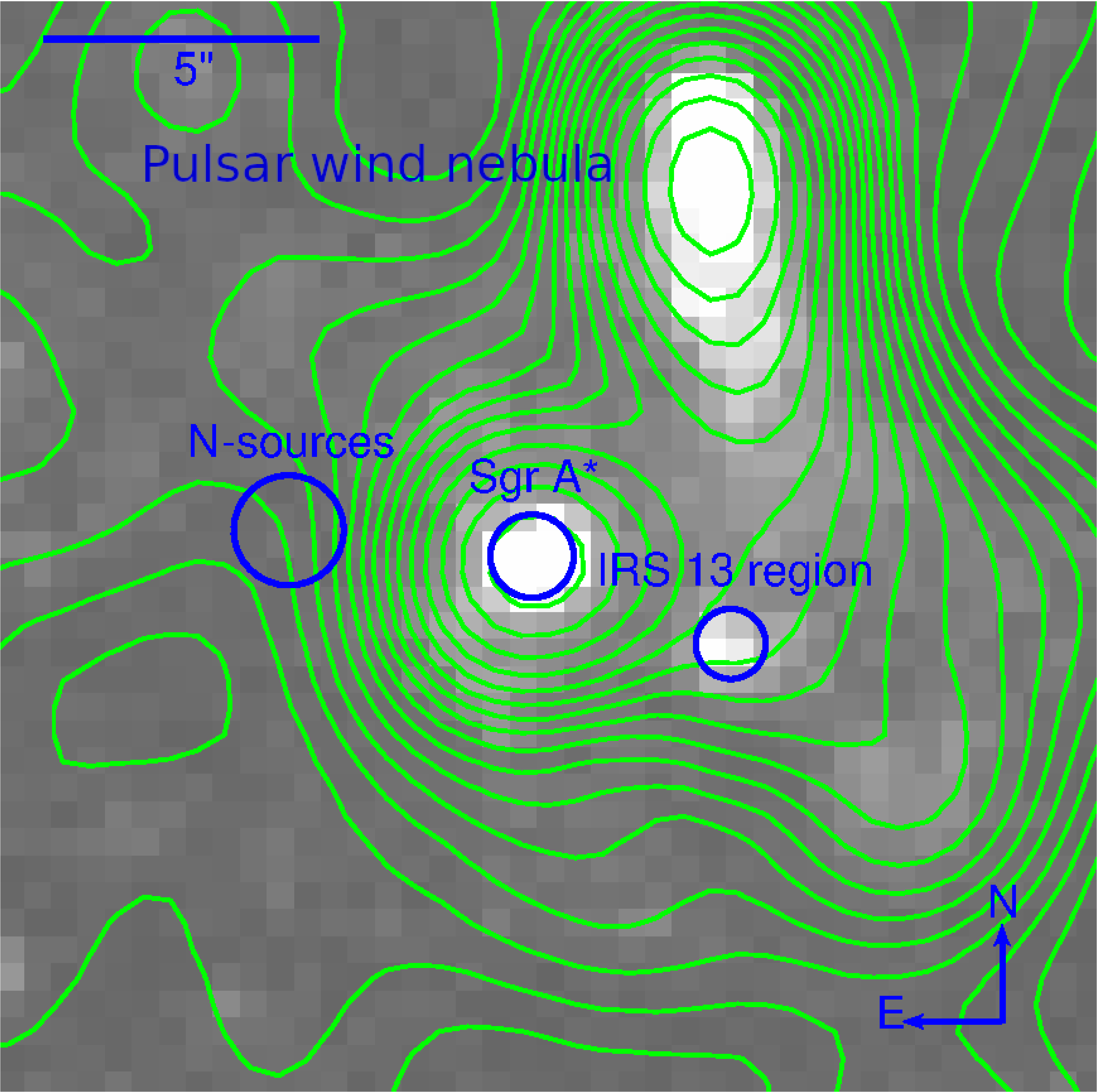}
   \caption{The \textit{Chandra} X-ray image from 4 to 8 keV. The north is up and the east is to the left. Distinct X-ray bright sources are labelled, namely Sgr~A*, IRS 13, and the pulsar wind nebula. There is no significant X-ray source at the position of the N-source association. The angular scale is indicated in the upper left corner. }
\label{Fig:x-ray}   
\end{center}
\end{figure}

Here we include additional plots and tables relevant for the study of the stellar association around IRS~1W. In Figure~\ref{fig:11 regions}, we depict 11 regions chosen
randomly to compare the number of stars inside them with the region around IRS~1W (N-source region). The proper motion of the N2 star is shown in Figure~\ref{plot:proper motion of w2}. In Figure~\ref{fig:Gaussian fit}, we illustrate that the NW-moving sources have a Gaussian-like distribution, which is demonstrated by a Gaussian fit. In Figure~\ref{Fig:x-ray}, we show 4-8 keV Chandra X-ray image of the Galactic center central parsec including the N-source region. Near-infrared magnitudes of the N-sources are listed in Table~\ref{table:Magnitudes}, while the colour indices are
summarized in Table~\ref{table:Color-color}. Table~\ref{table:Geometric centers} lists the detemined geometrical centers of the N-sources, NW-, and NE-moving groups.



\begin{table*}
\begin{center}
 \begin{tabular}{cccc}
 \hline
 \hline
Name  & $H$  & $K$ & $L$  \\
\hline
N1  &  12.05 $\pm$  0.03 &   9.18 $\pm$  0.02 &   6.19 $\pm$  0.05  \\
N2  &  14.62 $\pm$  0.79 &  12.26 $\pm$  0.29 &  10.45 $\pm$  0.10  \\
N3  &  15.03 $\pm$  0.03 &  13.43 $\pm$  0.32 &  10.46 $\pm$  0.21  \\
N4  &  16.03 $\pm$  0.38 &  14.57 $\pm$  0.61 &  11.73 $\pm$  0.32  \\
N5  &  17.10 $\pm$  1.17 &  14.71 $\pm$  0.33 &  11.98 $\pm$  0.35  \\
N6  &  15.73 $\pm$  0.45 &  13.32 $\pm$  0.03 &  11.65 $\pm$  0.18  \\
N7  &  15.50 $\pm$  0.16 &  15.01 $\pm$  0.85 &  12.77 $\pm$  0.44  \\
N8  &  15.28 $\pm$  0.55 &  12.79 $\pm$  0.22 &  11.21 $\pm$  0.19  \\
N9  &  12.81 $\pm$  0.03 &   9.94 $\pm$  0.07 &   9.06 $\pm$  0.02  \\
N10 &  13.41 $\pm$  0.18 &  11.27 $\pm$  0.04 &  10.19 $\pm$  1.15  \\
N11 &  14.17 $\pm$  0.49 &  11.64 $\pm$  0.15 &  10.57 $\pm$  1.01  \\
N12 &  15.79 $\pm$  0.04 &  13.21 $\pm$  0.03 &  10.36 $\pm$  0.37  \\
N13 &  16.38 $\pm$  1.61 &  13.90 $\pm$  0.43 &  10.68 $\pm$  0.25  \\
N14 &  15.33 $\pm$  0.31 &  14.22 $\pm$  1.45 &  10.46 $\pm$  0.22  \\
N15 &  14.73 $\pm$  0.32 &  12.94 $\pm$  0.26 &  10.10 $\pm$  0.24  \\
N16 &  15.41 $\pm$  0.18 &  13.55 $\pm$  0.68 &  10.29 $\pm$  0.29  \\
N17 &  15.43 $\pm$  1.35 &  13.07 $\pm$  0.97 &  10.28 $\pm$  0.30  \\
N18 &  13.47 $\pm$  0.15 &  10.98 $\pm$  0.21 &   8.55 $\pm$  0.21  \\
N19 &  12.44 $\pm$  0.05 &  10.07 $\pm$  0.13 &   9.08 $\pm$  0.13  \\
N20 &  17.01 $\pm$  0.53 &  14.24 $\pm$  0.17 &  13.49 $\pm$  0.79  \\
N21 &  17.04 $\pm$  0.69 &  15.21 $\pm$  1.54 &  12.87 $\pm$  0.44  \\
N22 &  18.36 $\pm$  0.65 &  14.96 $\pm$  0.63 &  13.44 $\pm$  0.76  \\
N23 &  16.01 $\pm$  0.81 &  14.18 $\pm$  0.52 &  11.60 $\pm$  0.28  \\
N24 &  15.12 $\pm$  0.48 &  14.90 $\pm$  0.27 &  11.24 $\pm$  0.28  \\
N25 &  15.97 $\pm$  0.54 &  13.72 $\pm$  0.29 &  11.89 $\pm$  0.11  \\
N26 &  15.64 $\pm$  0.15 &  14.29 $\pm$  0.75 &  11.62 $\pm$  0.19  \\
N27 &  16.78 $\pm$  0.22 &  14.89 $\pm$  0.26 &  11.79 $\pm$  0.38  \\
N28 &  16.85 $\pm$  1.02 &  14.12 $\pm$  0.31 &  12.47 $\pm$  0.22  \\
N29 &  16.19 $\pm$  1.35 &  14.36 $\pm$  1.79 &  12.47 $\pm$  0.22  \\
N30 &  16.01 $\pm$  0.11 &  13.53 $\pm$  0.08 &  12.26 $\pm$  0.30  \\
N31 &  18.33 $\pm$  1.95 &  15.52 $\pm$  0.15 &  13.23 $\pm$  0.28  \\
N32 &  17.46 $\pm$  0.35 &  14.82 $\pm$  0.12 &  13.35 $\pm$  0.13  \\
N33 &  16.64 $\pm$  0.36 &  14.80 $\pm$  0.09 &  12.86 $\pm$  0.29  \\
N34 &  19.22 $\pm$  1.77 &  16.43 $\pm$  0.37 &  13.95 $\pm$  1.24  \\
N35 &  16.79 $\pm$  0.08 &  15.19 $\pm$  0.28 &  13.15 $\pm$  0.30  \\
N36 &  12.26 $\pm$  0.02 &   9.71 $\pm$  0.01 &   8.93 $\pm$  0.03  \\
N37 &  16.15 $\pm$  0.05 &  15.14 $\pm$  2.39 &  13.75 $\pm$  0.20  \\
N38 &  15.78 $\pm$  0.15 &  13.10 $\pm$  0.04 &  12.82 $\pm$  0.37  \\
N39 &  17.06 $\pm$  0.72 &  14.61 $\pm$  0.27 &  14.90 $\pm$  0.23  \\
N40 &  16.40 $\pm$  0.43 &  14.28 $\pm$  0.33 &  13.55 $\pm$  0.08  \\
N41 &  16.71 $\pm$  1.87 &  14.52 $\pm$  0.22 &  13.58 $\pm$  0.06  \\
N42 &  15.41 $\pm$  1.01 &  13.02 $\pm$  0.44 &  12.35 $\pm$  0.08  \\
IRS 10W  &  12.74 $\pm$  0.08 &  10.11 $\pm$  0.06 &   6.99 $\pm$  0.03  \\
IRS 21   &  14.03 $\pm$  0.24 &  10.72 $\pm$  0.21 &   6.97 $\pm$  0.10  \\
IRS 16NW &  11.39 $\pm$  0.05 &   9.26 $\pm$  0.03 &   8.62 $\pm$  0.02  \\
  \hline
  \hline
 \end{tabular}
 \caption{Magnitudes of sources.}
 \label{table:Magnitudes}
\end{center}
\end{table*}

\begin{table*}
\begin{center}
 \begin{tabular}{ccc}
 \hline
 \hline
Name  & $H-K_s$  & $K_s-L'$   \\
\hline

N1 	& 	2.867 	$\pm$ 	0.375 	& 	2.984 	$\pm$ 	0.552 	\\
N2 	& 	2.366 	$\pm$ 	0.304 	& 	1.805 	$\pm$ 	0.84 	\\
N3 	& 	1.604 	$\pm$ 	0.386 	& 	2.964 	$\pm$ 	0.323 	\\
N4 	& 	1.467 	$\pm$ 	0.686 	& 	2.839 	$\pm$ 	0.712 	\\
N5 	& 	2.392 	$\pm$ 	0.481 	& 	2.737 	$\pm$ 	1.216 	\\
N6 	& 	2.406 	$\pm$ 	0.177 	& 	1.674 	$\pm$ 	0.452 	\\
N7 	& 	0.491 	$\pm$ 	0.957 	& 	2.233 	$\pm$ 	0.865 	\\
N8 	& 	2.486 	$\pm$ 	0.295 	& 	1.582 	$\pm$ 	0.592 	\\
N9 	& 	2.875 	$\pm$ 	0.069 	& 	0.877 	$\pm$ 	0.074 	\\
N10 	& 	2.146 	$\pm$ 	1.156 	& 	1.073 	$\pm$ 	0.183 	\\
N11 	& 	2.537 	$\pm$ 	1.021 	& 	1.066 	$\pm$ 	0.514 	\\
N12 	& 	2.575 	$\pm$ 	0.371 	& 	2.850 	$\pm$ 	0.050 	\\
N13 	& 	2.478 	$\pm$ 	0.500 	& 	3.226 	$\pm$ 	1.665 	\\
N14 	& 	1.107 	$\pm$ 	1.465 	& 	3.763 	$\pm$ 	1.480 	\\
N15 	& 	1.791 	$\pm$ 	0.354 	& 	2.836 	$\pm$ 	0.409 	\\
N16 	& 	1.865 	$\pm$ 	0.736 	& 	3.253 	$\pm$ 	0.702 	\\
N17 	& 	2.362 	$\pm$ 	1.014 	& 	2.782 	$\pm$ 	1.659 	\\
N18 	& 	2.482 	$\pm$ 	0.295 	& 	2.431 	$\pm$ 	0.255 	\\
N19 	& 	2.369 	$\pm$ 	0.182 	& 	0.983 	$\pm$ 	0.138 	\\
N20 	& 	2.775 	$\pm$ 	0.805 	& 	0.751 	$\pm$ 	0.556 	\\
N21 	& 	1.829 	$\pm$ 	1.604 	& 	2.333 	$\pm$ 	1.689 	\\
N22 	& 	3.406 	$\pm$ 	0.992 	& 	1.518 	$\pm$ 	0.909 	\\
N23 	& 	1.832 	$\pm$ 	0.592 	& 	2.588 	$\pm$ 	0.964 	\\
N24 	& 	0.219 	$\pm$ 	0.393 	& 	3.660 	$\pm$ 	0.552 	\\
N25 	& 	2.246 	$\pm$ 	0.315 	& 	1.828 	$\pm$ 	0.613 	\\
N26 	& 	1.348 	$\pm$ 	0.773 	& 	2.669 	$\pm$ 	0.763 	\\
N27 	& 	1.892 	$\pm$ 	0.459 	& 	3.099 	$\pm$ 	0.343 	\\
N28 	& 	2.723 	$\pm$ 	0.386 	& 	1.652 	$\pm$ 	1.071 	\\
N29 	& 	1.833 	$\pm$ 	1.802 	& 	1.885 	$\pm$ 	2.242 	\\
N30 	& 	2.475 	$\pm$ 	0.305 	& 	1.268 	$\pm$ 	0.130 	\\
N31 	& 	2.807 	$\pm$ 	0.317 	& 	2.291 	$\pm$ 	1.953 	\\
N32 	& 	2.638 	$\pm$ 	0.180 	& 	1.478 	$\pm$ 	0.373 	\\
N33 	& 	1.839 	$\pm$ 	0.303 	& 	1.934 	$\pm$ 	0.366 	\\
N34 	& 	2.795 	$\pm$ 	1.291 	& 	2.474 	$\pm$ 	1.810 	\\
N35 	& 	1.601 	$\pm$ 	0.450 	& 	2.041 	$\pm$ 	1.102 	\\
N36 	& 	2.545 	$\pm$ 	0.014 	& 	0.779 	$\pm$ 	0.022 	\\
N37 	& 	1.601 	$\pm$ 	0.409 	& 	2.041 	$\pm$ 	0.291 	\\
N38 	& 	1.008 	$\pm$ 	2.400 	& 	1.386 	$\pm$ 	2.392 	\\
N39 	& 	2.686 	$\pm$ 	0.375 	& 	0.275 	$\pm$ 	0.151 	\\
N40 	& 	2.447 	$\pm$ 	0.357 	& 	-0.284 	$\pm$ 	0.770 	\\
N41 	& 	2.118 	$\pm$ 	0.339 	& 	0.735 	$\pm$ 	0.543 	\\
N42 	& 	2.191 	$\pm$ 	0.232 	& 	0.94 	$\pm$ 	3.480 	\\
IRS 10W 	& 	2.626 	$\pm$ 	0.063 	& 	3.122 	$\pm$ 	0.096 	\\
IRS 21 	& 	3.307 	$\pm$ 	0.235 	& 	3.748 	$\pm$ 	0.323 	\\
IRS 16NW 	& 	2.132 	$\pm$ 	0.036 	& 	0.638 	$\pm$ 	0.057 	\\

  \hline
  \hline
 \end{tabular}
 \caption{$H-K_s$ and $K_s-L'$ colour indices for the N-sources.}
 \label{table:Color-color}
\end{center} 
\end{table*}



\begin{table}[tbh!]
\centering
 \begin{tabular}{ c cc}
 \hline
 \hline
Name  &  $\Delta\alpha$($arcsec$)& $\Delta\delta$ ($arcsec$) \\
 \hline
Geometric center of N sources (42)           & 6.24 $\pm$ 0.47     &  0.25 $\pm$	0.42 \\     
Geometric center of NW-flying sources (28)   & 6.21 $\pm$ 0.48     &	 0.35 $\pm$	0.42 \\ 
Geometric center of NE-flying sources (14)   & 6.38 $\pm$ 0.38     &	 0.10 $\pm$	0.39  \\     
  \hline
  \hline
 \end{tabular}
 \caption{Geometric centers of N sources, NW-flying sources, and NE-flying sources calculated as the medians of the corresponding values in Table~\ref{table:proper motion wsources km/s}. The geometric center of N sources (total) is in between NW-flying and NE-flying sources. It closer to NW-flying sources since there are twice as numerous as NE-flying sources. The uncertainty of the median value is calculated as the median of the measurement offsets from the median.}
 \label{table:Geometric centers}
\end{table}




\end{document}